\documentclass[citeautoscript,floatfix,aps,prl,twocolumn,superscriptaddress,longbibliography]{revtex4-1}  %
\usepackage{amsmath}
\usepackage{amssymb}
\usepackage{amsthm}

\usepackage{amsfonts}
\usepackage{graphicx}
\usepackage{hyperref}
\usepackage{array}
\usepackage{braket}
\usepackage{booktabs}
\usepackage{xcolor}
\usepackage{subcaption}
\usepackage{dsfont} 
\usepackage{xr}
\usepackage{mathtools}
\usepackage[export]{adjustbox}
\usepackage{tikz}
\usepackage{tensor}
\usepackage{tikz-network}
\usepackage{paratensor}
\usepackage{paraKDHlattice}

\newcommand{\id}{\mathrm{id}}

\newcommand{\C}{\mathbb{C}}
\newcommand{\calC}{\mathcal{C}}

\newcommand{\Rep}{\mathrm{Rep}}
\newcommand{\Tr}{\mathrm{Tr}}
\newcommand{\Hil}{\mathcal{H}}

\newcommand{\oA}{{o}}
\newcommand{\oB}{{s}}

\begin{document}
\title{Parastatistics and a secret communication challenge}
	\author{Zhiyuan Wang}
\affiliation{Max-Planck-Institut f{\"{u}}r Quantenoptik, Hans-Kopfermann-Str. 1, 85748 Garching, Germany}
	\date{\today}

\begin{abstract}
One of the most unconventional features of topological phases of matter is the emergence of quasiparticles with exotic statistics, such as non-Abelian anyons~\cite{MR1991,kitaev2003fault,kitaev2006anyons,Nayak2008NAAnyons} in two dimensional systems. Recently, a different type of exotic particle statistics that is consistently defined in any dimension, called $R$-parastatistics, %
is also shown to be possible in a special family of topological phases~\cite{wang2023para}. %
However, the physical significance of emergent parastatistics still remains elusive. 
Here we demonstrate a distinctive physical consequence of parastatistics by proposing a challenge game that can only be won using physical systems hosting paraparticles, 
as passing the challenge requires the two participating players to secretly communicate in an indirect way by exploiting the nontrivial exchange statistics of the quasiparticles. 
The winning strategy using emergent paraparticles is robust against noise, as well as the most relevant class of eavesdropping via local measurements. 
This provides both an operational definition and an experimental identity test for paraparticles, alongside a potential application in secret communication. 
\end{abstract}
	\maketitle
\paragraph{Introduction}
Topological phases~\cite{TPorder1,kitaev2003fault,levin2005string,kitaev2006anyons,Chen2010LUT,Wen2017Zoo,Barkeshli2019SETclassification} are long-range-entangled phases of matter, exhibiting fundamentally new physical properties compared to conventional symmetry breaking phases, %
and have attracted tremendous interest in the condensed matter community. A key unconventional feature of such phases is the emergence of exotic quasiparticle statistics beyond the celebrated Bose-Einstein and Fermi-Dirac statistics, which reflects the pattern of long-range entanglement~\cite{LevinWen2006TEE,KitaevPreskill2006TEE,Oshikawa2012quasiparticle} in the topologically ordered ground states. For example, non-Abelian anyonic statistics is possible in two dimensions~(2D)~\cite{MR1991,kitaev2003fault,kitaev2006anyons,Nayak2008NAAnyons}, 
where spatially braiding anyons leads to a unitary evolution in the topologically degenerate states with $n$ identical anyons, realizing a representation of the braid group $B_n$, and has a potential application in topological quantum computation~\cite{kitaev2003fault,Nayak2008NAAnyons}. %

In higher spatial dimensions, it has long been believed that all point-like particles are either fermions or bosons~\cite{doplicher1971local,doplicher1974local,LanKongWen3DAB,LanWen3DEF}. There is actually a straightforward generalization of non-Abelian statistics to arbitrary spatial dimension, known as parastatistics~\cite{Green1952}, in which identical particles transform in higher dimensional representations of the symmetric group $S_n$~(instead of the braid group $B_n$) under exchange of their spatial positions~\cite{Taylor1970b}. %
Although parastatistics has been considered since the dawn of quantum mechanics~\cite{nobelphysics1946} and extensively studied by the high energy physics community in the second half of the last century~\cite{Araki1961,Greenberg1965,LANDSHOFF196772,druhl1970parastatistics,Taylor1970b}, it is now widely believed to be physically equivalent to ordinary particle statistics~\cite{Greenberg1965,druhl1970parastatistics,doplicher1971local,doplicher1974local}. 
In other words, the theory of parastatistics, while mathematically consistent, was long thought to predict no new physical phenomena beyond what is already known from the theory of fermions and bosons.

Recently, a paper~\cite{wang2023para}  proposed a different theory of parastatistics, now called $R$-parastatistics, which exhibit exchange behaviors dramatically different from our conventional physical picture of fermions and bosons. These $R$-paraparticles have been shown to emerge in a family of exactly solvable quantum spin models~\cite{wang2023para} realizing an exotic class of topological phases. Despite this significant progress, many important  questions are open, such as what are the most distinctive physical~(experimental) consequence of $R$-paraparticles, and how to characterize their universal topological properties independently of the microscopic realization. %

In this paper we demonstrate a distinctive physical consequence of $R$-parastatistics from a quantum information viewpoint. Roughly speaking, we show that emergent $R$-parastatistics enable long-range communication of information by only exchanging the spatial positions of paraparticles, without either perturbing the environment, or leaving any trace of information behind. We formulate this idea by proposing a secret communication challenge game, designed in a way that can only be won by topological phases hosting emergent paraparticles~\footnote{Throughout this paper, whenever we claim that the game can only be won by emergent paraparticles, we refer to the 3D version of the game, since the 2D version can also be won by a special class of non-Abelian anyons,  and additional challenges need to be introduced to single out paraparticles in 2D, as we discuss in Sec.~\ref{SI:AAtwist} of the SM~\cite{Suppl}.}, as winning the challenge requires the two participating  players to send a message to each other by exploiting the nontrivial exchange statistics of the quasiparticles. 
We illustrate the winning strategy using exactly solvable quantum spin models with emergent $R$-paraparticles~(in both two and three dimensions), and show that this strategy is robust against local noise and eavesdropping.  %
We give a simple description of the universal topological properties of $R$-paraparticles that involves only basic quantum mechanics, and show that the winning strategy only relies on these universal properties.  
We then systematically investigate which three dimensional~(3D) topological phases can win the game, using the established fact that point-like topological quasiparticles in 3D are universally described by symmetric fusion categories~(SFCs)~\cite{doplicher1971local,*doplicher1974local,LanKongWen3DAB,LanWen3DEF}.
This leads to a categorical description of $R$-paraparticles and the winning strategy, and singles out a special class of 3D topological phases that can win the game. 

\paragraph{Rules of the game}
The participants of the game consist of two parties: two players, who we call Alice~(A) and Bob~(B), and a group of Referees~(R). %
To win the game, the two players are required to send a message to each other using a very restricted class of local operations on a common quantum many body system, 
and the Referees' role is to initiate the challenge and monitor the whole process of the game to prevent cheating.

\begin{figure}
		\centering
\begin{subfigure}[t]{.95\linewidth}
	\centering\includegraphics[width=\linewidth]{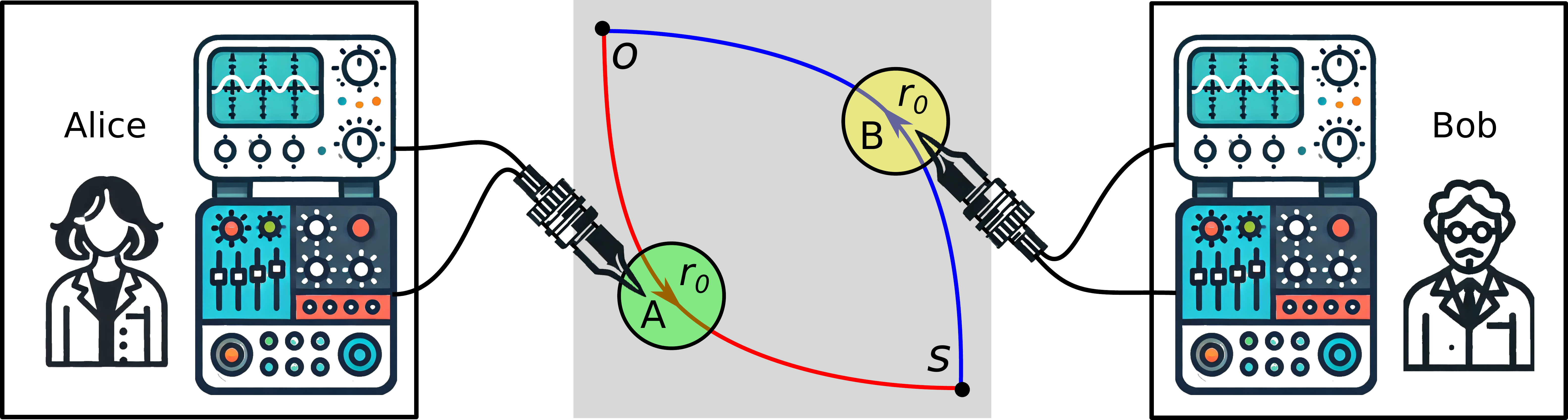}
	\caption{\label{fig:ABcontrol} Local operations controlled by the players}%
\end{subfigure}
\begin{subfigure}[t]{.324\linewidth}
	\centering\includegraphics[width=.96\linewidth]{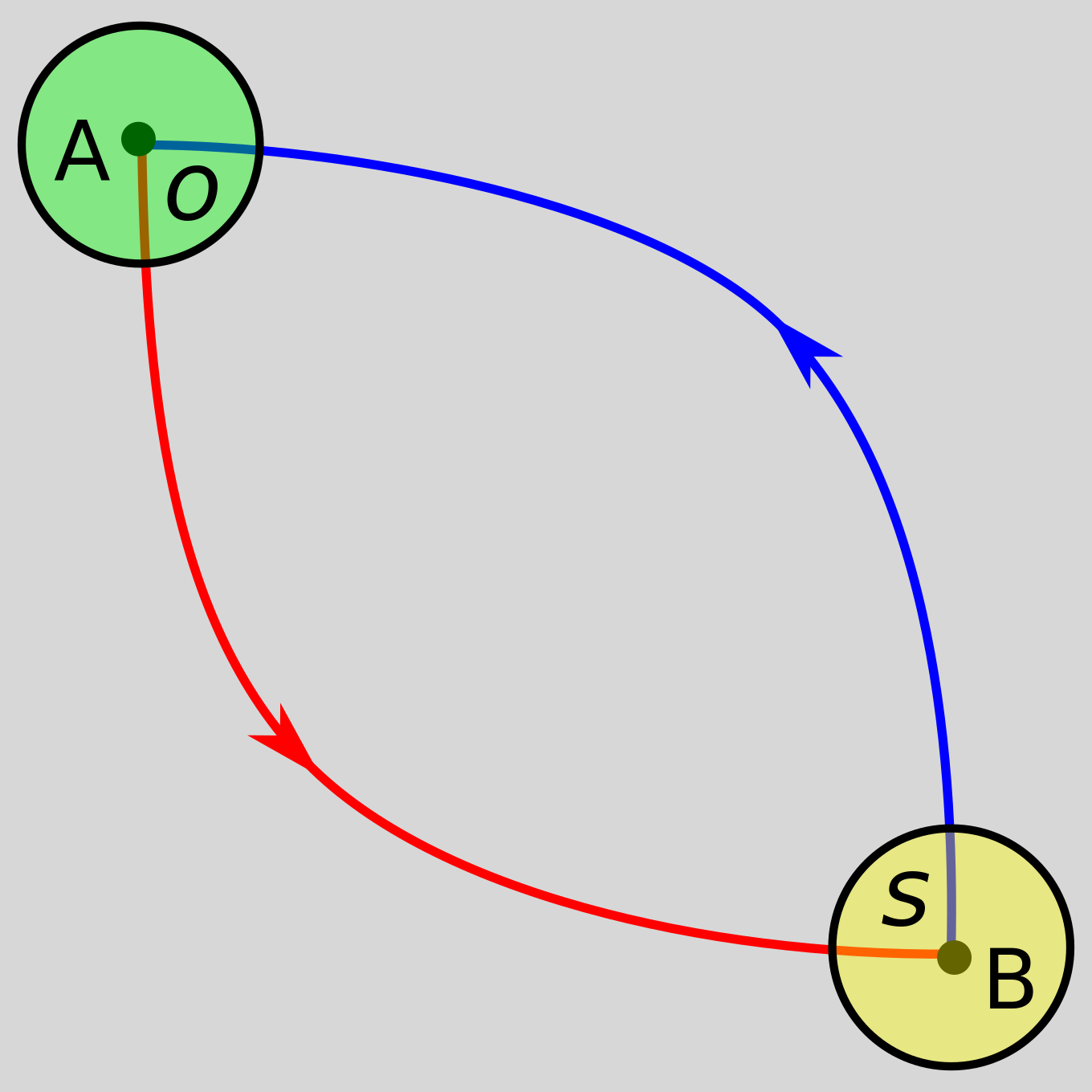} 
	\caption{\label{fig:gamestart} Start $t=0$}
\end{subfigure}
\begin{subfigure}[t]{.324\linewidth}
	\centering\includegraphics[width=.96\linewidth]{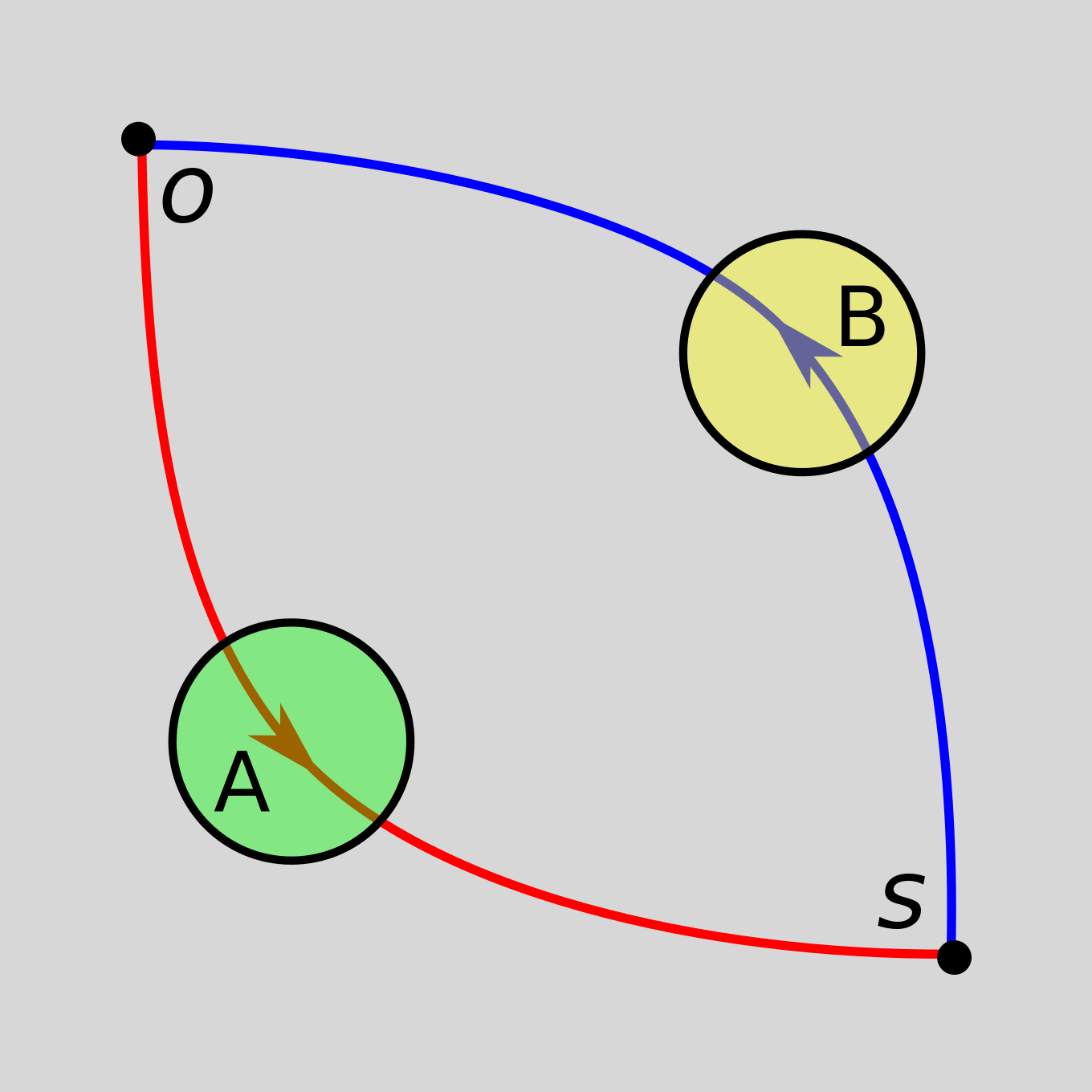} 
	\caption{\label{fig:duringgame} $0<t<T$}
\end{subfigure}
\begin{subfigure}[t]{.324\linewidth}
	\centering\includegraphics[width=.96\linewidth]{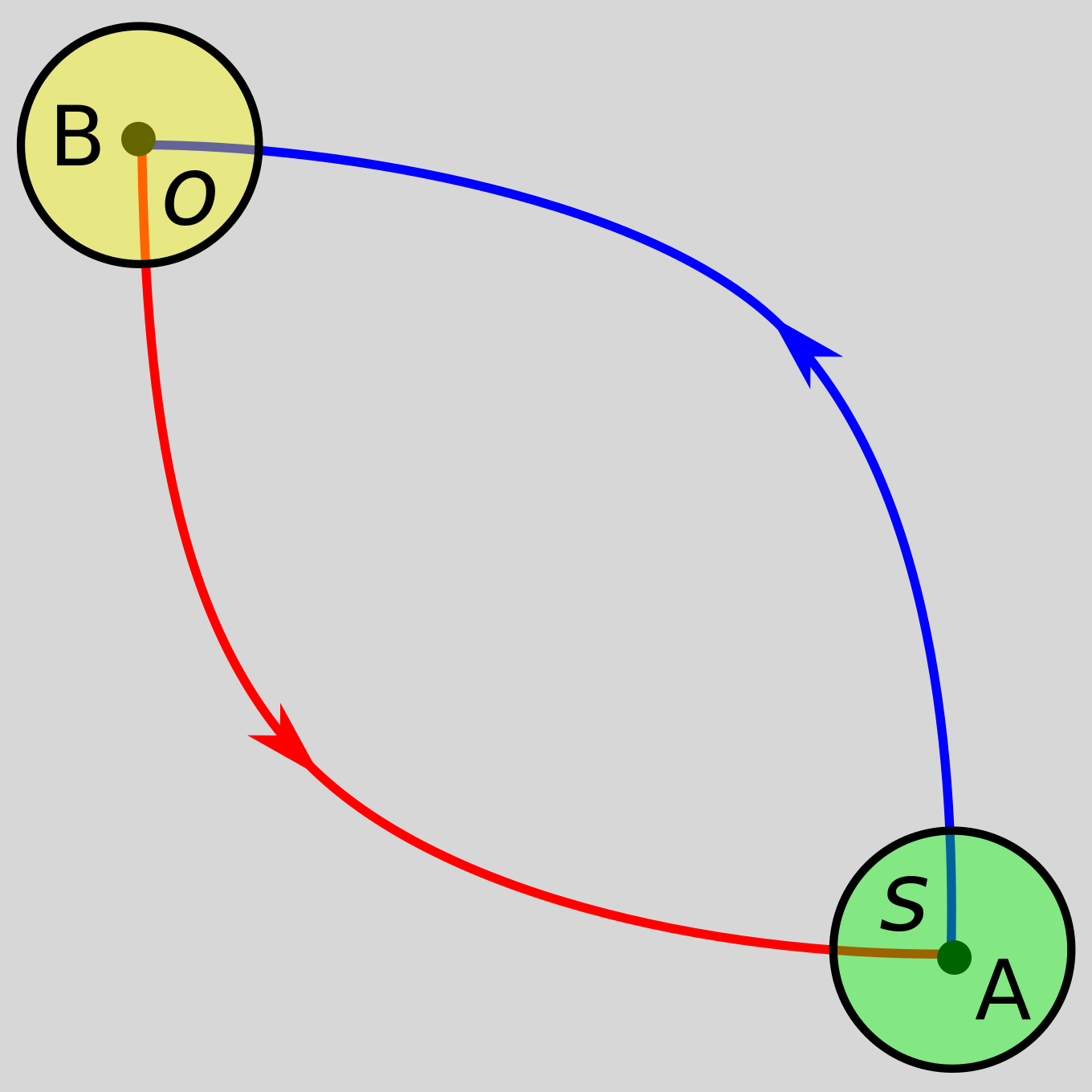} 
	\caption{\label{fig:gameend} End $t=T$}
\end{subfigure}
\caption{\label{fig:game} Illustration of the game process. %
Here we draw the 2D version for simplicity; in 3D, the circles become spheres. 
The two circles have radius $r_0$ defined by the players. During the game the Referees move the two circles along their respective paths to complete an exchange of positions. The two far separated points $\oA,\oB$ are chosen by the players, while the two paths are chosen by the Referees. %
} 
\end{figure}

Before the game starts, the players are allowed to discuss a winning strategy, and then they are required to\\
(1) submit a locally-interacting  Hamiltonian $\hat{H}$ on a 2D or 3D lattice %
that has a unique, gapped, and frustration-free ground state $\ket{G}$;\\ %
(2) choose the radius $r_0$ of the circles~(or spheres in 3D) in Fig.~\ref{fig:game}, %
and two far separated~\footnote{For example, we can require that the distance between $\oA$ and $\oB$ is at least half of the system size. } points $\oA,\oB$ in the lattice~(which are allowed to be on the boundary);\\ %
(3) experimentally prepare the ground state $\ket{G}$ for a sufficiently large system size $L\gg r_0$~[$L$ is determined by the Referees after they receive (1) and (2)].\\
A few technical remarks are in order here. %
Gapped means that there is a uniform lower bound~(independent of the system size $L$) on the energy difference between the first excited state and ground state of $\hat{H}$. 
Frustration-free means that $\hat{H}$ can be written in the form $\hat{H}=\sum_{i}\hat{h}_i$ such that $\hat{h}_i\geq 0$ and $\hat{h}_i\ket{G}=0$,
which is a technical requirement imposed to simplify discussion, and we  relax this requirement in the supplemental material~(SM)~\cite{Suppl}.  %
A few other technical regularity conditions are also stated in the SM~\cite{Suppl}.

When the Referees receive all above (1)-(3), they %
 experimentally check that $\ket{G}$ is prepared correctly, %
by verifying that $\hat{h}_i\ket{G}=0$ %
for all $i$. Then Alice and Bob are led to separate rooms, and the Referees randomly pick two numbers $a,b\in\{1,2,3,4\}$, and give them to Alice and Bob, respectively, who do not know each other's number. 
The goal for the players is to gain information about each other's number through a restricted set of local operations on the physical system they have prepared, as we describe below.

Once the game starts, any form of direct communication between the players are prohibited. Throughout the game, Alice and Bob are confined in separate rooms, %
where the only access to the outside world is through
local unitary operations and measurements on the system they prepared, restricted to their respective circle areas, as shown in Fig.~\ref{fig:ABcontrol}. %
When the game starts at $t=0$, the state of the physical system is initialized to be $\ket{\Psi(0)}=\ket{G}$, %
and Alice's circle starts at the special point $\oA$, while Bob's circle starts at the other special point $\oB$, as shown in Fig.~\ref{fig:gamestart}. 
Then the Referees select the two paths~\cite{RefereeChoosePath} shown in Fig.~\ref{fig:game}, and slowly move the two circles along their respective paths simultaneously, ensuring that the two circles remain far apart at all times. 
Throughout the process, the Referees frequently check the local ground state condition $\hat{h}_i\ket{\Psi(t)}=0$ %
everywhere beyond the two circle areas %
to make sure that the players are not cheating by leaving any trace of information behind. If at any moment, the Referees detect an excitation beyond the circle areas, the challenge fails. %
The game ends at $t=T$ in the configuration shown in Fig.~\ref{fig:gameend}, when the two circles complete an exchange of positions. After this, the Referees move both circles out of the system and perform one last check of the local ground state condition $\hat{h}_i\ket{\Psi(T)}=0$
everywhere. 
If this final check is passed, the Referees ask Alice about $b$, and ask Bob about $a$, and the players win if they both answer correctly.

\begin{figure}
	\begin{subfigure}[t]{.32\linewidth}
		\centering\includegraphics[width=.9\linewidth]{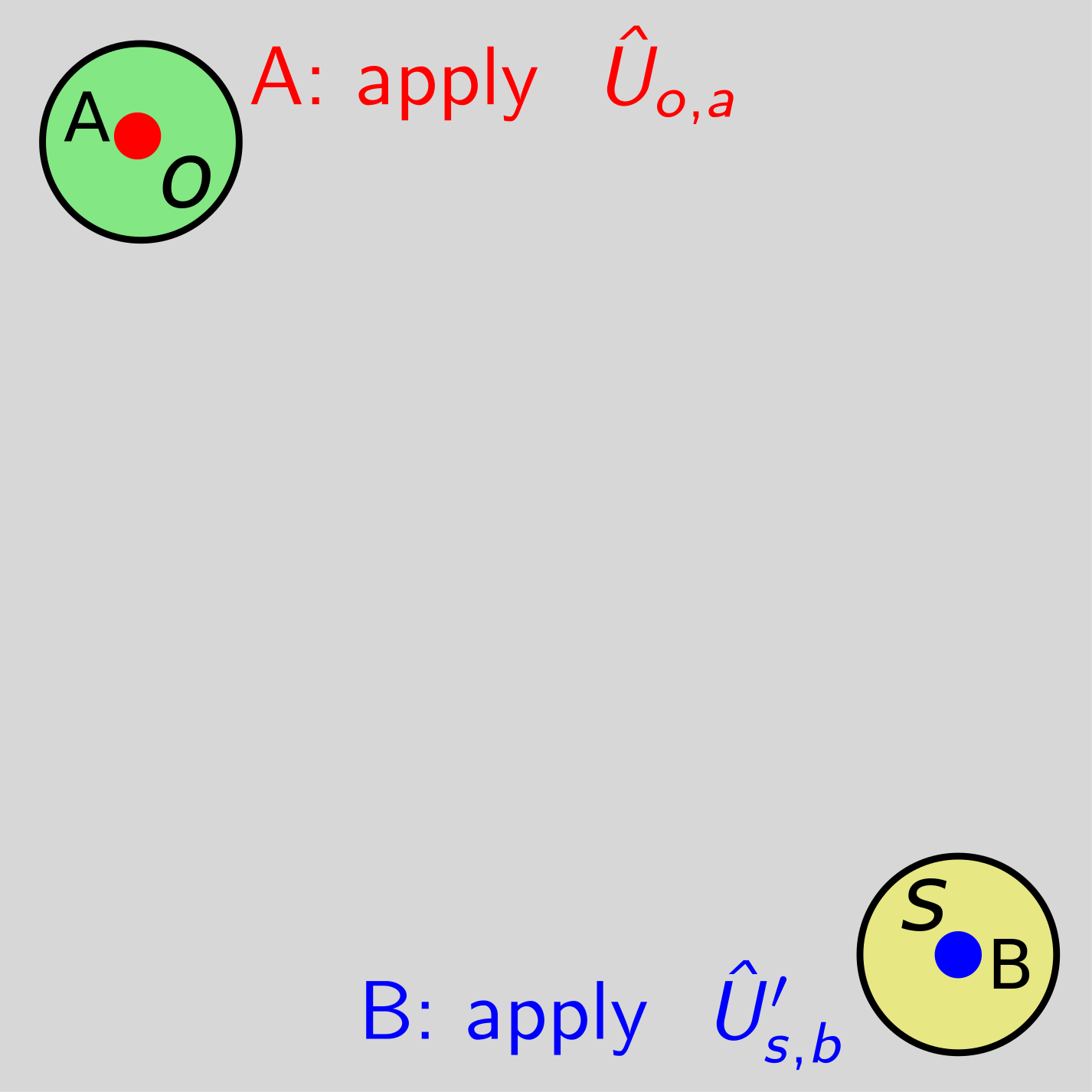} 
		\caption{\label{fig:WSstart} Start $t=0$}
	\end{subfigure}
	\begin{subfigure}[t]{.32\linewidth}
		\centering\includegraphics[width=.9\linewidth]{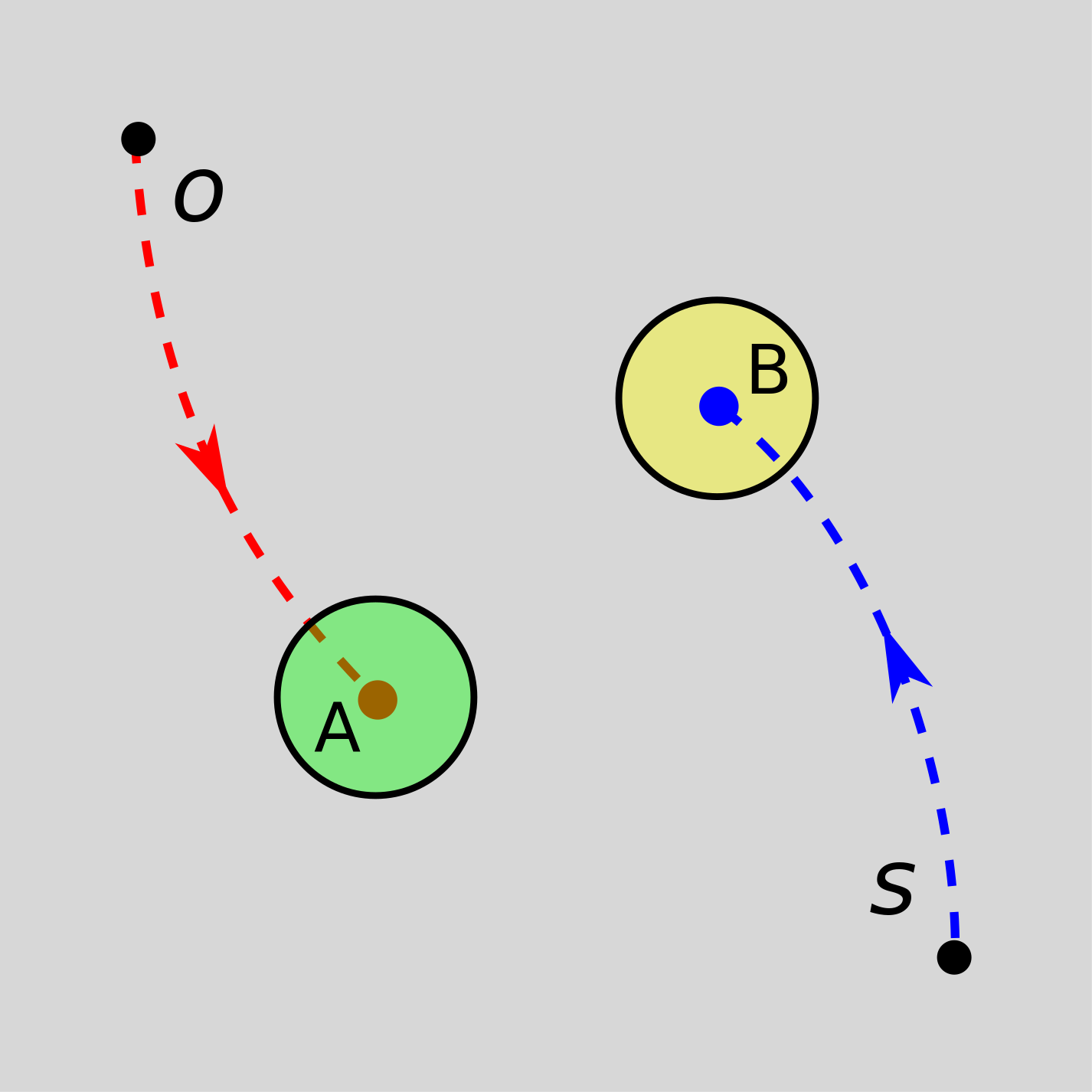} 
		\caption{\label{fig:WSduringgame} $0<t<T$}
	\end{subfigure}
	\begin{subfigure}[t]{.32\linewidth}
		\centering\includegraphics[width=.9\linewidth]{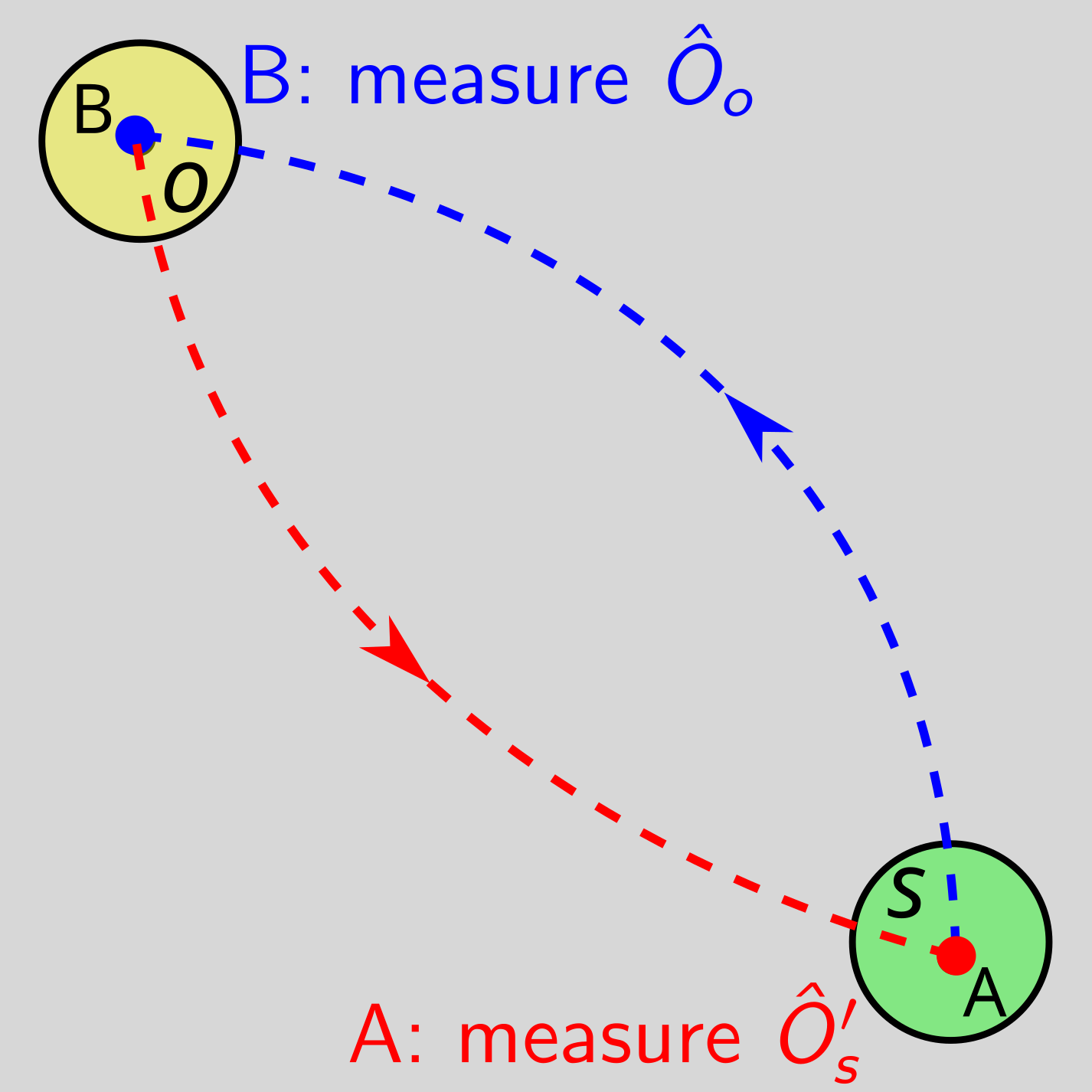} 
		\caption{\label{fig:WSgameend} End $t=T$}
	\end{subfigure}
	\caption{\label{fig:winstrategy} Illustration of the winning strategy using emergent paraparticles. $\oA$ and $\oB$ are chosen to be the two special points 
	where a paraparticle can be locally created and measured, as described in properties (5) and (6).  The dashed lines represent the paths traversed by the circles, along which the players have applied unitary operators $\hat{U}_{ij}$ in Eq.~\eqref{eq:paraparticlemove}. 
	}
\end{figure}

What can the players do to win the game? They can
create some excitations inside their respective circle areas and make measurements on them, but throughout the experiment they are obliged to move the excitations to follow the circle movement, and clean up whatever excitations inside the circles before the game ends at $t=T$, to make sure that the Referees cannot detect an excitation anywhere beyond the circle areas at any time. %
At first glance, the challenge appears insurmountable under such restrictive conditions, as the players need to communicate some information without either leaving a trace behind, or sending any physical particles to each other as information carriers. %
Nevertheless, we will see %
that it is possible to succeed %
if the ground state $\ket{G}$ hosts some emergent quasiparticles with non-trivial exchange statistics. The players can create such quasiparticles at $t=0$, and encode their numbers in the internal state of the quasiparticles. Then they move the quasiparticles following the circle movement, and right before the game ends at $t=T$, the quasiparticles have exchanged their positions, inducing a unitary rotation on their internal states, which the players can measure. We will see that some information can be sent this way if the exchange statistics is nontrivial. 
In the following, we illustrate this winning strategy in detail, using emergent paraparticles in the some exactly solvable models. 
\paragraph{Solvable quantum spin models with emergent $R$-paraparticles} can be constructed in both 2D and 3D, as we present explicitly in the SM~\cite{Suppl}.
Both models have unique, gapped, and frustration-free ground state $\ket{G}$ and therefore satisfy the game requirement. The microscopic details of these models are not important, as it turns out that the winning strategy depends only on certain universal topological properties of the emergent paraparticles that are common to both models, which we summarize below:\\ 
(1) The state space. In general, an excited state of $\hat{H}$ with $n$ identical paraparticles is denoted as $\ket{G;i_1^{a_1} i_2^{a_2} \ldots i_n^{a_n}}$. 
Here $i_1,i_2,\ldots, i_n$ are the positions of the paraparticles, and we assume that they are mutually different~\footnote{If not, then the exclusion statistics of the paraparticles have to be taken into account~\cite{wang2023para}. This is unnecessary for our discussion here, since we never move two paraparticles to the same position in this paper.}. %
The numbers $a_1,a_2,\ldots,a_n\in \{1,2,\ldots,m\}$ label the \textit{internal states} of the paraparticles, where $m\geq 2$ is an integer called the quantum dimension of the paraparticle. When the positions $i_1,\ldots, i_n$ are fixed, there are $m^n$ linearly independent internal states, spanning a $m^n$-dimensional subspace which we denote as $\Hil_{i_1i_2 \ldots i_n}$. \\
(2) Topological degeneracy. The internal states of the paraparticles cannot be accessed using any local measurements when all particles are far away from the boundary and from each other. Formally, this means that any local operator $\hat{O}$ is diagonal in the subspace $\Hil_{i_1 i_2 \ldots i_n}$
\begin{equation}\label{eq:TPdegenerate}
\!\braket{G;i_1^{b_1} \ldots i_n^{b_n}|\hat{O}|G;i_1^{a_1} \ldots i_n^{a_n}}=C^O_{i_1\ldots i_n}\!\prod_{j=1}^n \delta_{a_j b_j}, %
\end{equation}
where $C^O_{i_1\ldots i_n}$ is a constant. Eq.~\eqref{eq:TPdegenerate} implies that states in $\Hil_{i_1 i_2 \ldots i_n}$ are topologically degenerate in energy~\footnote{In a general topological phase, Eq.~\eqref{eq:TPdegenerate} only holds up to a small correction of order $O(e^{-l/\xi})$~\cite{hastings2005quasiadiabatic1}, where $\xi$ is the correlation length of $\ket{G}$ and $l$ is the minimal distance between the paraparticles~(and the boundary). Consequently, the topological degeneracy in energy generally also has a small splitting up to $O(e^{-l/\xi})$. In this solvable model, however, we can tune the model parameters such that $\xi=0$, and this small correction exactly vanishes. }.\\
(3) Particle movements. %
There exists a local unitary operator $\hat{U}_{i_k j_k}$ that moves the particle at position $i_k$ to a nearby position $j_k$, for any $k=1,2,\ldots,n$:
\begin{equation}\label{eq:paraparticlemove}
\hat{U}_{i_k j_k}\ket{G;i_1^{a_1}\ldots i_k^{a_k}\ldots i_n^{a_n}}=\ket{G;i_1^{a_1}\ldots j_k^{a_k} \ldots i_n^{a_n}}.
\end{equation}
(4) Exchange statistics. 
Here we focus on the case $n=2$ which is enough for our purpose; the generalization to arbitrary number of paraparticles is straightforward~\cite{Suppl}. For  $i\neq j$, 
$\Hil_{i j}$ must be equal to $\Hil_{j i}$, since both describe the subspace of excited states of two identical paraparticles at positions $i$ and $j$. Therefore the two different basis $\{\ket{G;i^{a}j^b}|1\leq a,b\leq m\}_{}$ and $\{\ket{G;j^{b}i^{a}}|1\leq a,b\leq m\}_{}$ must be related by a unitary transformation:
\begin{equation}\label{eq:exchangestatR}
	\ket{G;i^a j^b}=\sum_{a',b'} R^{b'a'}_{ab}\ket{G;j^{b'} i^{a'}},
\end{equation}
where the four-index tensor $R^{b'a'}_{ab}$~(called the $R$-matrix) must satisfy the Yang-Baxter equation~[Eq.~\eqref{eq:YBE} in the SM~\cite{Suppl}] so that Eq.~\eqref{eq:exchangestatR} can be consistently generalized to any number of particles~\cite{wang2023para}. 
Each $R$-matrix defines a type of parastatistics~\cite{wang2023para}.   
There are many different types of parastatistics that can emerge in condensed matter systems, %
and for the two solvable models we present in the SM~\cite{Suppl}, 
the relevant paraparticles have quantum dimension $m=4$, and the nonzero elements of the $R$-matrices are given by
\begin{equation}\label{eqApp:seth-R}
R^{b'a'}_{ab}=\pm 1, \text{ if }	(b',a')=
	\left(
	\begin{array}{cccc}
		43 & 12 & 24 & 31 \\
		21 & 34 & 42 & 13 \\
		14 & 41 & 33 & 22 \\
		32 & 23 & 11 & 44 \\
	\end{array}
	\right)_{ab},%
\end{equation}
where $(4,3)$ is abbreviated as $43$ and similarly for others, and $-1$~($+1$) is for the 2D~(3D) model.\\
(5) Creation and annihilation of paraparticles. %
Paraparticles can be locally created at the two special points $\oA,\oB$ shown in Fig.~\ref{fig:winstrategy}. More precisely, there exists unitary operators $\hat{U}_{\oA,a}$, $\hat{U}'_{\oB,a}$ localized around $\oA,\oB$, respectively, satisfying
\begin{eqnarray}\label{eq:localcreationatcorner}
\hat{U}_{\oA,a}\ket{G;i_1^{a_1} \ldots i_n^{a_n}}&=&\ket{G;\oA^a i_1^{a_1} \ldots i_n^{a_n}},\nonumber\\
\hat{U}'_{\oB,a}\ket{G;i_1^{a_1} \ldots i_n^{a_n}}&=&\ket{G;i_1^{a_1} \ldots i_n^{a_n}\oB^a}.
\end{eqnarray}
Since unitary processes are reversible, one can also annihilate paraparticles at $\oA,\oB$ using $\hat{U}^\dagger_{\oA,a}$, $\hat{U}^{\prime\dagger}_{\oB,a}$, respectively.\\
(6) Measurement of the internal state. When a paraparticle is close to either $\oA$ or $\oB$, one can measure its internal state locally: there exist observables $\hat{O}_{\oA}$, $\hat{O}'_{\oB}$ localized around $\oA,\oB$, respectively, satisfying
\begin{eqnarray}\label{eq:localmeasurementatcorner}
	\hat{O}_{\oA}\ket{G;i_1^{a_1} \ldots i_n^{a_n}}&=&a_1 \ket{G; i_1^{a_1} \ldots i_n^{a_n}},\text{ if } i_1=\oA,\nonumber\\
	\hat{O}'_{\oB}\ket{G;i_1^{a_1} \ldots i_n^{a_n}}&=&a_n \ket{G;i_1^{a_1} \ldots i_n^{a_n}},\text{ if } i_n=\oB.
\end{eqnarray}
Note that if we instead have $i_k=\oA$ for some $k>1$~(or $i_k=\oB$ for some $k<n$), then we need to use a basis transformation in Eq.~\eqref{eq:exchangestatR}~[or more generally, Eq.~\eqref{eq:exchangestatR-npt}] %
to swap $i_k$ all the way to the front~(back) before applying Eq.~\eqref{eq:localmeasurementatcorner} to compute %
$\braket{\hat{O}_{\oA}}$~(or $\braket{\hat{O}'_{\oB}}$). We will see an example soon in Eq.~\eqref{eq:Psi3}.  %

All these properties can be derived from the exact solution of these two models, as we show in the SM~\cite{Suppl}. We emphasize that in order to win the challenge, it is crucial to have two special points $\oA$ and $\oB$ in the lattice where a single paraparticle can be created and measured using local operations, and these are the two points the players choose at which they begin and end their journey.
In the 2D model, 
$\oA$ and $\oB$ are chosen to be 
at the intersection between two different types of gapped boundaries, while in the 3D model, they are chosen to be the positions of some special point-like topological defects in the bulk. %

\paragraph{Winning strategy with emergent paraparticles}
We now detail the strategy that ensures success in the challenge. 
The pregame preparation is done using either of the two solvable models mentioned above: 
the players submit the model Hamiltonian $\hat{H}$, choose $\oA,\oB$ to be the two special points in Eqs.~(\ref{eq:localcreationatcorner},\ref{eq:localmeasurementatcorner}), 
prepare its ground state $\ket{G}$, 
and let the Referees verify. %
After the game starts and the players obtain their numbers from the Referees, they use the following strategy to win, as illustrated in Fig.~\ref{fig:winstrategy}:\\ 
(a) At $t=0$, Alice applies $\hat{U}_{\oA,a}$ to create a paraparticle with internal state $a$ at $\oA$, and 
similarly Bob applies $\hat{U}'_{\oB,b}$ at  $\oB$.  %
The state of the whole system is %
$\ket{\Psi(0)}=\ket{G;\oA^a \oB^b}$;\\
(b) Throughout the game process, %
Alice and Bob move their paraparticles along the paths using $\hat{U}_{ij}$ in Eq.~\eqref{eq:paraparticlemove},
closely following the circle movement;\\
(c) When the exchange is complete, the state evolves to
\begin{eqnarray}\label{eq:Psi3}
	\ket{\Psi(T)}=\ket{G;\oB^a \oA^b}=\sum_{a',b'}R^{b'a'}_{ab}\ket{G;\oA^{b'} \oB^{a'}},
\end{eqnarray}
where we use Eq.~\eqref{eq:exchangestatR}. Then Alice and Bob measure $\hat{O}'_{\oB}$ and $\hat{O}_{\oA}$ in Eq.~\eqref{eq:localmeasurementatcorner} %
and obtain $a'$ and $b'$, respectively, after which the state collapses into $\ket{G;\oA^{b'} \oB^{a'}}$. Finally, Alice and Bob annihilate their paraparticles using $\hat{U}^{\prime\dagger}_{\oB,a'}$ and $\hat{U}^{\dagger}_{\oA,b'}$, respectively, so that the Referees cannot detect any excitations anywhere after they leave the game. 

According to the $R$-matrix in Eq.~\eqref{eqApp:seth-R}, %
the measurement results $a',b'$ are definite, and importantly,
knowing $b$ and $b'$ completely determines $a$, and similarly, knowing $a$ and $a'$ completely determines $b$~\footnote{Indeed, this $R$-matrix is an example of a perfect tensor, i.e. if we group any two indices of the $R$-matrix as input, and the other two as output, the resulting matrix we get is unitary~(it is an invertible classical gate in our case). This allows the players to obtain complete information about each other's number by inverting this matrix.}. For example, if Alice has $(a,a')=(2,3)$, then she searches in the second row of the matrix for the column that has $a'=3$. This turns out to be the fourth column, so she determines that $b=4$ and $b'=1$.  %
This allows the players to win the game with a 100\% success rate.

This winning strategy demonstrates a  clear physical distinction between paraparticles and ordinary fermions or bosons with an internal symmetry, such as color or flavor. The statistics of the latter can also be described %
by Eq.~\eqref{eq:exchangestatR}, but with
$R^{b'a'}_{ab}=\pm\delta_{a'a}\delta_{b'b}$. 
With such an $R$-matrix, Alice will simply obtain $a'=a$ and Bob will obtain $b'=b$, which contains no information about each other's number, and therefore cannot win the game with a probability better than pure guessing. A more careful argument that ordinary fermions and boson cannot win the game is given in the SM~\cite{Suppl}, and a formal proof is given in Ref.~\cite{wang2025secret}.

 It is important to emphasize again that this strategy relies only on the topological properties (1)-(6) of emergent paraparticles, which are universal characteristics of the underlying topological phase independent of the detailed microscopic realization. In particular, any quantum system hosting quasiparticles satisfying properties (1)-(6) with a non-trivial $R$-matrix~(i.e., $R^{b'a'}_{ab}$ is not of the trivial product form $R^{b'a'}_{ab}=p_{a'a}q_{b'b}$) can pass the challenge. Therefore, we can use properties (1)-(6) as a definition of $R$-paraparticles~\cite{wang2025secret}\footnote{We mention that operational axioms motivated by quantum information theory for defining exotic particle statistics have also been explored in Ref.~\cite{Dakic2024reconstructionof}.}.

Finally, while winning the challenge in 3D unambiguously demonstrates non-trivial $R$-parastatistics~\cite{RefereeChoosePath}, in 2D, it is possible that some special types of non-Abelian anyons can also win. 
In the SM~\cite{Suppl} we design a twisted version of the game that can prevent a class of non-Abelian anyons from winning, and more general anti-anyon twists are given in Ref.~\cite{wang2025secret}. 
\paragraph{Robustness against noise and eavesdropping}
Importantly, the above winning strategy using emergent paraparticles is robust against noise, 
which is a consequence of Eq.~\eqref{eq:TPdegenerate}. 
More precisely, the susceptibility to local noise decays exponentially as $e^{-l/\xi}$, where $l$ is the minimal distance between the paraparticles and the boundary. This means that 
even when noise is present, the above  strategy still has a high chance of success that is independent of the system size $L$. 
Eq.~\eqref{eq:TPdegenerate} also implies the robustness against the most physically relevant class of eavesdropping via local measurements: when both players are in the bulk, a thief cannot obtain any information about the players' numbers $a$ and $b$ using any local measurements on the system, even including measurements at $\oA,\oB$ or inside the circle areas. %

\paragraph{Symmetric fusion category description}
So far we have presented solvable models in 2D and 3D with emergent paraparticles that can pass the challenge. 
We now give an alternative description of the winning strategy 
using the framework of SFC, which gives a universal description of point-like quasiparticles in 3D topological phases~\cite{doplicher1971local,*doplicher1974local,kong2014braided,Zhu3DTPOModels,LanKongWen3DAB,LanWen3DEF,Johnson-Freyd2022}. We will see that certain 3D topological phases described by a special class of SFCs can win the game, an example is the 3D solvable model given in the SM~\cite{Suppl}. %
The SFC description of the winning strategy is based on the following space-time diagram:
\begin{equation}\label{eq:SFCdescriptiongame}
\adjincludegraphics[height=14ex,valign=c]{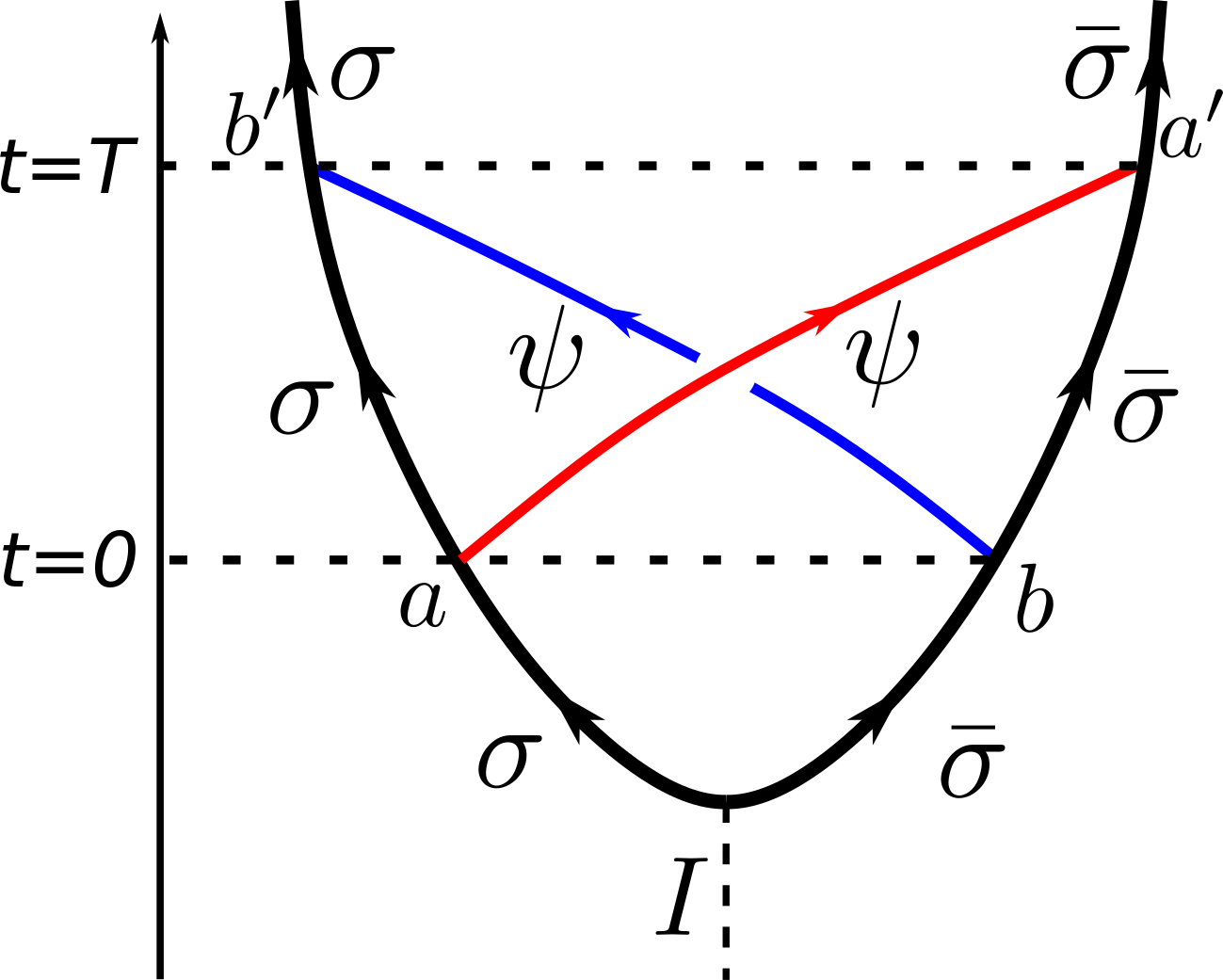}
=R^{b'a'}_{ab}~\adjincludegraphics[height=14ex,valign=c]{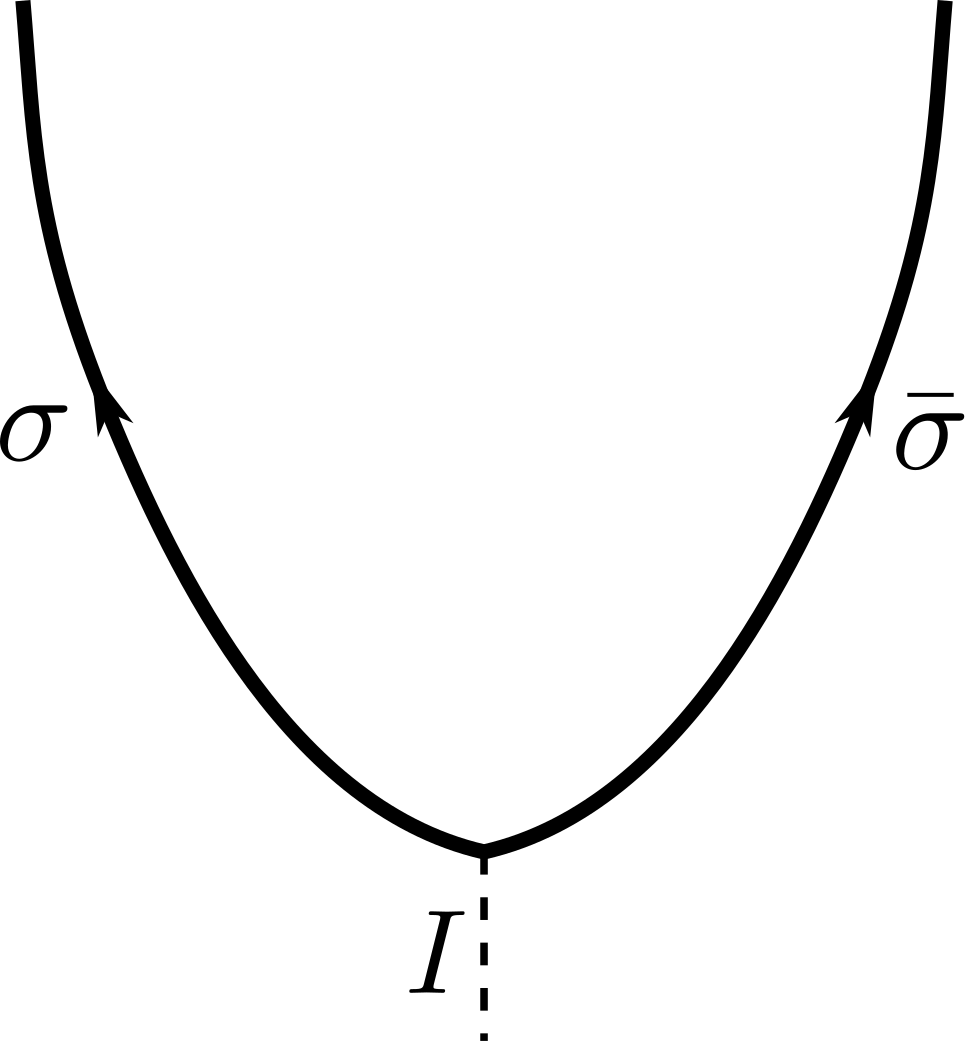},
\end{equation}
where $\sigma$ represents a special type of topological quasiparticle, with $\bar{\sigma}$ being its antiparticle. %
In this case the players need to prepare a state with topological quasiparticles $\sigma$ and $\bar{\sigma}$  at positions $\oA$ and $\oB$, respectively, which can still be the unique and gapped ground state of a locally-interacting~(but not translationally invariant) Hamiltonian that qualifies the requirement of the game, as shown in the SM~\cite{Suppl}.  
Importantly, we assume that the quasiparticles have fusion rules of the form~\footnote{Indeed, it is straightforward to show that the first one in Eq.~\eqref{eq:sigmapsifusion} implies the second.}
\begin{equation}\label{eq:sigmapsifusion}
	\sigma \times \psi=m~\sigma,\quad   \psi\times \bar{\sigma}=m~\bar{\sigma},
\end{equation}
where $m$ is the quantum dimension of $\psi$. Eq.~\eqref{eq:sigmapsifusion} implies that $\psi$ can be locally created and measured in the vicinity of $\sigma$ or $\bar{\sigma}$, %
and the tensor $R^{b'a'}_{ab}$ in Eq.~\eqref{eq:SFCdescriptiongame} %
defines the statistics of the paraparticle $\psi$ in the sense of Eq.~\eqref{eq:exchangestatR}.  
Indeed, Eq.~\eqref{eq:SFCdescriptiongame} exactly describes the winning strategy similar to that depicted in Fig.~\ref{fig:winstrategy}, in which at $t=0$ Alice creates a paraparticle with internal state $a$ in the vicinity of $\sigma$, and meanwhile Bob creates a paraparticle with internal state $b$ in the vicinity of $\bar{\sigma}$; at $t=T$ when the exchange process is over, %
Alice measures the internal state of her paraparticle in the vicinity of $\bar{\sigma}$ and obtains $a'$, and then fuses her paraparticle into $\bar{\sigma}$, and meanwhile Bob performs similar operations in the vicinity of $\sigma$ and obtains $b'$.  %
As long as $R$ is nontrivial, some information can be transferred between Alice and Bob, and by using multiple identical layers of the same system, the players can transfer more information to each other, thereby arbitrarily enhancing the chance of winning. 
We therefore conclude that a gapped ground state $\ket{G}$ described by an SFC $\mathcal{C}$ can win the challenge if $\mathcal{C}$ has fusion rules of the form~\eqref{eq:sigmapsifusion} such that the $R$-matrix defined by Eq.~\eqref{eq:SFCdescriptiongame} is nontrivial. 
The 3D solvable model given in the SM~\cite{Suppl} is described by an SFC of this type. 
A more detailed categorical analysis %
is presented in a separate paper~\cite{wang2025secret}. 

\paragraph{Discussion}
We have presented a secret communication challenge that demonstrates a distinctive physical feature of paraparticles. %
In a nutshell, emergent parastatistics in topological phases enables long-range communication between the two players solely through an exchange of their positions, without either perturbing the environment, or leaving any trace of information behind. Such a winning strategy using parastatistics is robust against any local noise and eavesdropping.  %
We can therefore use this challenge game  as a criterion and an operational definition for non-trivial parastatistics. %

We emphasize that our secret communication challenge games fundamentally differ from prior non-local games in topological phases~\cite{Burnell2023TCnonlocalgame,Nandkishore2025TCnonlocalgame} in goal and design. 
These prior non-local games are played on the 2D toric code, their main goal is to demonstrate noise-robust quantum advantage; while our goal is to demonstrate the nontrivial physical consequence of $R$-parastatistics, an intrinsically quantum many-body phenomenon that only occur in certain exotic topological phases. %

Finally, we discuss some potential future directions. 
One interesting direction is to experimentally implement this challenge game and its winning strategy using systems that host emergent paraparticles. The first critical step would be  engineering such a system and preparing its ground state.
The latter can potentially be achieved by generalizing the recent topological quantum state preparation protocol~\cite{HierarchyTPO_LOCC,Iqbal2024}  involving measurements and feedforward to quantum double ground states based on an exotic class of solvable groups~\cite{Howlett1982}. 
Another interesting direction is to generalize the challenge game and its winning strategy to quantum many body systems at finite temperature~(and more generally to mixed states).
In 2D, there is no topological order at finite temperature~\cite{Dennis2002TQM,Castelnovo2007Entanglement,Nussinov2009PNAS,Hastings2011finiteTTPO}. In 3D, however, some nontrivial features of topological order persist at finite temperature%
~\cite{TCFiniteT2008,Cheng2025finiteTTPO}. 
Successfully achieving this challenge at finite temperatures is crucial for experimentally observing emergent parastatistics.
Additionally, this could highlight a nontrivial feature of finite-temperature topological order and its potential application in quantum information. 

\acknowledgments
We thank Alexei Kitaev, Xiao-Gang Wen, J. Ignacio Cirac, Meng Cheng,  Kaden Hazzard, Norbert Schuch, Chong Wang, Dominic Else, Sung-Sik Lee, Tao Shi, and Xiaoqi Sun for discussions. This work is supported by the Munich Quantum Valley~(MQV), which is supported by the Bavarian state government with funds from the Hightech Agenda Bayern Plus, and 
the Munich Center for Quantum Science and Technology~(MCQST), funded by the Deutsche Forschungsgemeinschaft~(DFG) under Germany's Excellence Strategy~(EXC2111-390814868).

\bibliography{CGE,/home/lagrenge/Dropbox/zoteroLibrary/library1}

\begin{thebibliography}{77}
\bibitem [{Ref()}]{RefereeChoosePath}%
  \BibitemOpen
  \href@noop {} {}\bibinfo {note} {We emphasize that, to establish the claim
  that only paraparticles can win the challenge in 3D, it is very important
  that we ask the Referees to choose the two paths in Fig.~\ref{fig:game}. If
  we instead allow the players to choose their paths, then they can win the
  game using non-Abelian anyons even in 3D. For example, they can construct a
  3D system by stacking identical layers of some 2D topological phase hosting
  non-Abelian anyons, and then choose the two paths to lie entirely in the same
  layer, which reduces the quantum dynamics to a 2D problem, allowing them to
  use a 2D winning strategy with non-Abelian anyons. Asking the Referees to
  choose the paths (after the players submit the physical system) forces the
  players to use excitations with unrestricted mobility in 3D, thereby
  preventing the above strategy with non-Abelian anyons (whose mobility is
  necessarily restricted to some 2D surface).}\BibitemShut {Stop}%
\end{thebibliography}

\clearpage 

\begin{center}
	\textbf{\large Supplementary Information for ``Parastatistics and a secret communication challenge''}
\end{center}
\makeatletter
\def\renewtheorem#1{%
	\expandafter\let\csname#1\endcsname\relax
	\expandafter\let\csname c@#1\endcsname\relax
	\gdef\renewtheorem@envname{#1}
	\renewtheorem@secpar
}
\def\renewtheorem@secpar{\@ifnextchar[{\renewtheorem@numberedlike}{\renewtheorem@nonumberedlike}}
\def\renewtheorem@numberedlike[#1]#2{\newtheorem{\renewtheorem@envname}[#1]{#2}}
\def\renewtheorem@nonumberedlike#1{  
	\def\renewtheorem@caption{#1}
	\edef\renewtheorem@nowithin{\noexpand\newtheorem{\renewtheorem@envname}{\renewtheorem@caption}}
	\renewtheorem@thirdpar
}
\def\renewtheorem@thirdpar{\@ifnextchar[{\renewtheorem@within}{\renewtheorem@nowithin}}
\def\renewtheorem@within[#1]{\renewtheorem@nowithin[#1]}
\makeatother

\setcounter{equation}{0}
\setcounter{page}{1}
\setcounter{section}{0}

\setcounter{secnumdepth}{1}
\setcounter{secnumdepth}{2}
\setcounter{secnumdepth}{3}
\makeatletter
\renewtheorem{theorem}{Theorem}[section]
\renewcommand{\theequation}{S\arabic{equation}}
\renewcommand{\thefigure}{S\arabic{figure}}

\renewcommand{\bibnumfmt}[1]{[S#1]}
\renewcommand{\thetable}{S\arabic{table}}
\renewcommand{\thesection}{S\arabic{section}}

This supplemental material fills in several technical details omitted in the main text. 
Specifically, in Sec.~\ref{sec:technical_requirement_H} we list out all the technical requirements on the Hamiltonian $\hat{H}$ of the physical system and present a version of the game that relaxes the frustration-free condition. 
In Secs.~\ref{sec:2DESMreview} and \ref{SI:3Dmodelparaparticle} we provide more microscopic details about the 2D and the 3D exactly solvable models and show how to extract the universal properties of emergent paraparticles from their exact solution.
In Sec.~\ref{SI:AAtwist} we present a simple example of anti-anyon twist that prevents a class of non-Abelian anyons from passing the 2D version of the challenge. 
In Sec.~\ref{sec:noise_robust} we remark in greater detail on the robustness of the winning strategy against local noise and eavesdropping. 
In Sec.~\ref{sec:Cat_analysis} we present a sketch of the argument that only nontrivial paraparticles can pass the 3D version of the challenge.

\section{Technical requirements on the Hamiltonian $\hat{H}$}\label{sec:technical_requirement_H}
In the following we give all the technical requirements on the Hamiltonian $\hat{H}$ of the physical system that the players are allowed to use to win the challenge. 
Essentially, we want to formulate a regularity condition on $\hat{H}$ so that it has a well-defined thermodynamic limit and describes a gapped phase of matter. 
Formally, the requirements on $\hat{H}$ are:\\
(a) $\hat{H}$ is required to be translationally invariant except at the two special points $\oA$ and $\oB$ chosen by the players, i.e., 
$\hat{H}$ should be of the form $\hat{H}=\sum_i\hat{h}_i$, where $\hat{h}_i$ and $\hat{h}_j$ should be related by a translation unless one of $i$ and $j$ is close to either $\oA$ or $\oB$;\\
(b) $\hat{H}$ is allowed to have an open boundary condition, which can be freely chosen by the players; however, we require that the boundary can be divided into at most $d$ simply connected regions~(where $d$ is the spatial dimension), each region $S$ is described by a specific boundary type~[i.e. the boundary Hamiltonian in $S$ has the form $\sum_{i\in S}\hat{h}'_i$, where $\hat{h}'_i$ is allowed to differ from the bulk Hamiltonian $\hat{h}_i$, but for any $i,j\in S$~(with $i,j\notin\{\oA,\oB\}$), $\hat{h}'_i$ and $\hat{h}'_j$ should be related by a translation].\\ %
(c) $\hat{H}$ has a unique, gapped, and frustration-free ground state $\ket{G}$;\\

We emphasize that the frustration-free condition in (c) applies to the whole system, including at the boundary~(if any) and the special points $\oA,\oB$. For example, if open boundary condition is used, the condition (c) requires that both the  bulk Hamiltonian $\hat{h}_i$ and the boundary terms $\hat{h}'_i$ are positive semidefinite and annihilate the unique, gapped ground state  $\ket{G}$.

\subsection{Relaxing the frustration free condition}\label{sec:relax_FF}
As we mentioned earlier, the requirement that the ground state is frustration free can be relaxed. To this end we use an important theorem by Hastings~\cite{hastings2006solving} which claims that if a  locally interacting Hamiltonian $\hat{H}=\sum_i \hat{h}_i$ has a unique, gapped ground state $\ket{G}$, then %
 $\hat{H}$ can be rewritten as $\hat{H}=E_0+\sum_i \hat{h}'_i$, where $E_0$ is the ground state energy, and $\hat{h}'_i$ is localized around $i$ with  subexponentially decaying tails,  satisfying $\hat{h}'_i\ket{G}=0$ for all $i$. In this way, the Referees can still prevent cheating by measuring $\hat{h}'_i$ beyond the circle areas throughout the game, and the players can still manage to localize their quasiparticles inside the circle areas, if they choose $r_0$ to be far greater then the correlation length of the system. 

\section{A 2D solvable model with emergent paraparticles}\label{sec:2DESMreview}
In this section we briefly review the 2D exactly solvable model introduced in Ref.~\cite{wang2023para}, which host emergent $R$-paraparticles that can be used to win the challenge.  
We begin by presenting the model Hamiltonian in Sec.~\ref{sec:2DESMreview_Hamiltonian} and its free paraparticle solution in Sec.~\ref{sec:2DESM_solution}. 
Then in Sec.~\ref{sec:2DESM_OBC} we show that by choosing a suitable hybrid open boundary condition, one can locally create a single paraparticle at some special point-like defects on the boundary.  
In Sec.~\ref{sec:2DESM-universal} we show how to derive the universal properties of the emergent paraparticles~[Eqs.~(\ref{eq:TPdegenerate}-\ref{eq:localmeasurementatcorner}) in the main text] from the exact solution of this model. %
Finally, in Sec.~\ref{sec:cat_description} we briefly mention the categorical description of the paraparticles and the gapped phases of matter in this model.
\subsection{Review of the model}\label{sec:2DESMreview_Hamiltonian}
\begin{figure}
	\centering
	\begin{tikzpicture}[baseline={([yshift=.4ex]current bounding box.center)}, scale=0.89]
		\ShadedTCLattice{0}{0}{2}{0.23}
		\OBClattice{0}{0}{2}
		\node  at (9*\hL,-7*\hL) {{\color{blue}$p$}};
		\node  at (3*\hL,-9*\hL) {{\color{red}$\nu$}};
		\ParticlePath{{2/-8,4/-8,4/-10,2/-10,2/-8/no}}{red}
		\IsoDots{{3/-7,5/-9,3/-11,1/-9}}{red}
		\ParticlePath{{8/-6/no,8/-8,10/-8,10/-6,8/-6}}{blue}
		\IsoDots{{9/-5,11/-7,9/-9,7/-7}}{blue}
		\ParticlePath{{0/0,2/0,4/0,4/-2,4/-4}}{parapurple}
		\IsoDots{{1/-1,5/-1,3/-3}}{parapurple}%
		\StringLabels{{0/0/\oA/above left,2/0/2/above left,4/0/4/above left,4/-2/6/above left,4/-4/i/below left}}{parapurple}%
		\node [fill=white,inner sep =0.5pt,minimum height=8pt] at (1*\hL,-1*\hL) [above left=0.1cm] {\small {\color{parapurple}1}};
		\node [fill=white,inner sep =0.5pt,minimum height=8pt] at (5*\hL,-1*\hL) [above left=0.1cm] {\small {\color{parapurple}5}};
		\node [fill=white,inner sep =0.5pt,minimum height=8pt] at (3*\hL,-3*\hL) [above left=0.1cm] {\small {\color{parapurple}7}};
		\node  at (4*\hL,-4*\hL)[above right] {${\color{parapurple}\hat{\psi}^+_{i,a}}$};
		\node  at (3.5*\hL,-0.5*\hL) {\large ${\color{parapurple}\Gamma}$};
		\node  at (12*\hL,-12*\hL) [below right] {$\oB$};
	\end{tikzpicture}
	\caption{\label{fig:mainlattice} The 2D solvable spin model with open boundary conditions~(figure adapted from Ref.~\cite{wang2023para}). Each black dot represents a 16 dimensional qudit, %
		and each open circle represents a 64 dimensional qudit. %
		Each colored triangle represents a 3-body interaction between qudits on its 3 vertices. In addition, there is an eight-body interaction term around every gray plaquette $\nu$ and every white plaquette $p$.
		Emergent paraparticles are created by matrix product string operators, an example of such an operator $\hat{\psi}_{i,a}^+$ is given in 
		Eq.~\eqref{eq:JWT_string_2D}, which acts consecutively on all the purple dots on and near the purple string $\Gamma$.  %
	}
\end{figure}
The model is constructed on a 2D lattice as illustrated in Fig.~\ref{fig:mainlattice}.
The model Hamiltonian consists of two parts $\hat{H}=\hat{H}_1+\hat{H}_2$, where
\begin{eqnarray}\label{def:2DsolvableH}
	\hat{H}_1&=&\sum_{\nu}\hat{A}_\nu+\sum_p \hat{B}_p,\nonumber\\
	\hat{H}_2&=&-\sum_{\langle ij\rangle}\hat{h}_{ij}-\sum_{l,a} \mu_l \hat{y}^+_{l,a}\hat{y}^-_{l,a},
\end{eqnarray}
where $\nu,p$ run over all the shaded and  white plaquettes, respectively, $l$ runs over all black dots, $a$ runs over the set $\{1,2,3,4\}$, and $\langle ij\rangle$ runs over all neighboring pairs of black dots with each pair appearing exactly once. 
All the local interacting terms in Eq.~\eqref{def:2DsolvableH} are defined in the SI of Ref.~\cite{wang2023para} and the explicit data are given in the accompanying code of the same paper. Below we present a convenient tensor network representation of these terms but without defining the individual tensors. 

The term $\hat{h}_{ij}$ represents a three-body interaction between the vertices of the triangle containing the directed edge $\langle ij\rangle$,
defined as
\begin{eqnarray}\label{def:3bodyterm}
	\scalebox{0.8}{
		\begin{tikzpicture}[baseline={([yshift=-.5ex]current bounding box.center)}, scale=.75]
			\TriangleTermIllustration{0*\hL}{-0*\hL}
	\end{tikzpicture}}=
	J_{ij}
	\scalebox{0.8}{\begin{tikzpicture}[baseline={([yshift=.4ex]current bounding box.center)}, scale=.8]
			\xtriangle{0}{0}{-}{i}\umatrix{1}{0}{w^+}{}\ytriangle{2}{0}{+}{j}
			\node at (1,-0.5) [below] {\small $k$};
	\end{tikzpicture}}+\mathrm{h.c.}\equiv \hat{h}_{ij},%
\end{eqnarray}
where we use the standard tensor network notation in which a contracted vertex indicates a summation. Each tensor defines a set of local quantum operators, 
and the labels $i,j,k$ indicate the positions of the lattice sites on which the local operators act on. 
For example, 
$\braket{p|\hat{y}^\pm_{a}|q}\equiv\!\!\begin{tikzpicture}[baseline={([yshift=-.5ex]current bounding box.center)}, scale=0.7]
	\ytriangle{0}{0}{\pm}{}
	\node  at (-1.4*\AL,0) {\footnotesize $a$};
	\quantumindices{0}{0}{}{}
	\node  at (.4*\AL,.8*\AL) {\footnotesize $p$};
	\node  at (.4*\AL,-.8*\AL) {\footnotesize $q$};
\end{tikzpicture}$
defines a local operator acting on a black site, where $a=1,2,3,4$ and  $\ket{p},\ket{q}$ labels basis states of the local Hilbert space on a black site. Similarly,
$\hat{w}^+_{ab}=
\scalebox{1}{\begin{tikzpicture}[baseline={([yshift=-.8ex]current bounding box.center)}, scale=.8]
		\umatr{0}{0}{w^+}{}
		\paraindices{0}{0}{a}{b}	
\end{tikzpicture}}$ 
is a local operator %
acting on a white site, where $w$ is one of $u_L,u_R,v_L$, or $v_R$ depending on the type of triangle in the lattice. 

		The terms $\hat{A}_\nu,\hat{B}_p$ are eight-body interaction acting on gray and white plaquettes, respectively. They have the following matrix product operator representation~\footnote{
			One can check that the operators $\hat{A}_\nu$ and $\hat{B}_p$ defined in Eq.~\eqref{eq:def_AvBp} are already Hermitian, so there is no need to manually add the Hermitian conjugate in the RHS, as it was originally done in Ref.~\cite{wang2023para}.
		}
		\begin{eqnarray}\label{eq:def_AvBp}
			\!\!\!\!\!\!\!\scalebox{0.75}{
				\begin{tikzpicture}[baseline={([yshift=.4ex]current bounding box.center)}, scale=.85]
					\DiamTermIllustrationP{0}{0}
					\node at (0,0) {$\hat{B}_p$};
			\end{tikzpicture}}\!\!&=&\scalebox{0.8}{
				\begin{tikzpicture}[baseline={([yshift=.4ex]current bounding box.center)}, scale=.8]
					\umatrix{0}{0}{v_L^+}{1}\Tpmatrix{1}{0}{2}\umatrix{2}{0}{v_L^+}{3}\Tpmatrix{3}{0}{4}
					\umatrix{4}{0}{v_R^+}{5}\Tpmatrix{5}{0}{6}\umatrix{6}{0}{v_R^+}{7}\Tpmatrix{7}{0}{8}
					\PBCcontract{-0.5}{0}{7.5}{1.5}
			\end{tikzpicture}}~,\nonumber\\ %
			\!\!\!\!\!\!\!\scalebox{0.75}{
				\begin{tikzpicture}[baseline={([yshift=.4ex]current bounding box.center)}, scale=.85]
					\node at (0,0) {$\hat{A}_\nu$};
					\DiamTermIllustrationV{0*\hL}{-0*\hL}
			\end{tikzpicture}}\!\!&=&\scalebox{0.8}{
				\begin{tikzpicture}[baseline={([yshift=.4ex]current bounding box.center)}, scale=.8]
					\umatrix{0}{0}{u_R^+}{1}\Tpmatrix{1}{0}{2}\umatrix{2}{0}{u_R^+}{3}\Tpmatrix{3}{0}{4}
					\umatrix{4}{0}{u_L^+}{5}\Tpmatrix{5}{0}{6}\umatrix{6}{0}{u_L^+}{7}\Tpmatrix{7}{0}{8}
					\PBCcontract{-0.5}{0}{7.5}{1.5}
			\end{tikzpicture}}~,\nonumber\\ %
		\end{eqnarray}
		where $\hat{T}^+_{ab}=\scalebox{1}{\begin{tikzpicture}[baseline={([yshift=-.8ex]current bounding box.center)}, scale=.8]
				\Tpmatrix{0}{0}{}
				\paraindices{0}{0}{a}{b}
		\end{tikzpicture}}$ is a local operator acting on a  black site. 
		Note that with open boundary condition those loop terms lying on the boundary are truncated, where white circles outside the boundary are absent. In this case the tensors $u^+_L,u^+_R,v^+_L,v^+_R$ on the absent site is replaced by a $\delta$ tensor, i.e., $\hat{w}^+_{ab}=\delta_{ab}$, for $w=u_L,u_R,v_L,v_R$. We will explain the importance of the open boundary condition in greater detail in Sec.~\ref{sec:2DESM_OBC}. 
		
		The loop terms $\hat{A}_\nu,\hat{B}_p$ satisfy
		\begin{equation}
			[\hat{A}_\nu,\hat{B}_p]=[\hat{A}_\nu,\hat{h}_{ij}]=[\hat{B}_p,\hat{h}_{ij}]=0,
		\end{equation}
		for any $\nu,p$ and $\langle ij\rangle$, which can be shown straightforwardly using a tensor network derivation~\cite{wang2023para}. 
		Therefore, $\hat{A}_\nu,\hat{B}_p$ define local conserved quantities of this system.  
		These conservation laws decompose the full Hilbert space as a direct sum of subspaces, where each subspace consists of common eigenstates of all $\hat{A}_\nu,\hat{B}_p$ and is invariant under the Hamiltonian $\hat{H}$. 
		In this paper we focus on the subspace in which all $\hat{A}_\nu,\hat{B}_p$ have minimal eigenvalues, which we refer to as the zero-vortex sector denoted by $\Phi_0$. %
		This sector contains the ground state $\ket{G}$ of $\hat{H}$, and within this sector the model can be exactly mapped to a system of free paraparticles, which we present in the following. 
		
		\subsection{The exact mapping to free paraparticles}\label{sec:2DESM_solution}
\begin{figure}
			\begin{subfigure}[t]{.38\linewidth}
				\includegraphics[width=.75\linewidth,left]{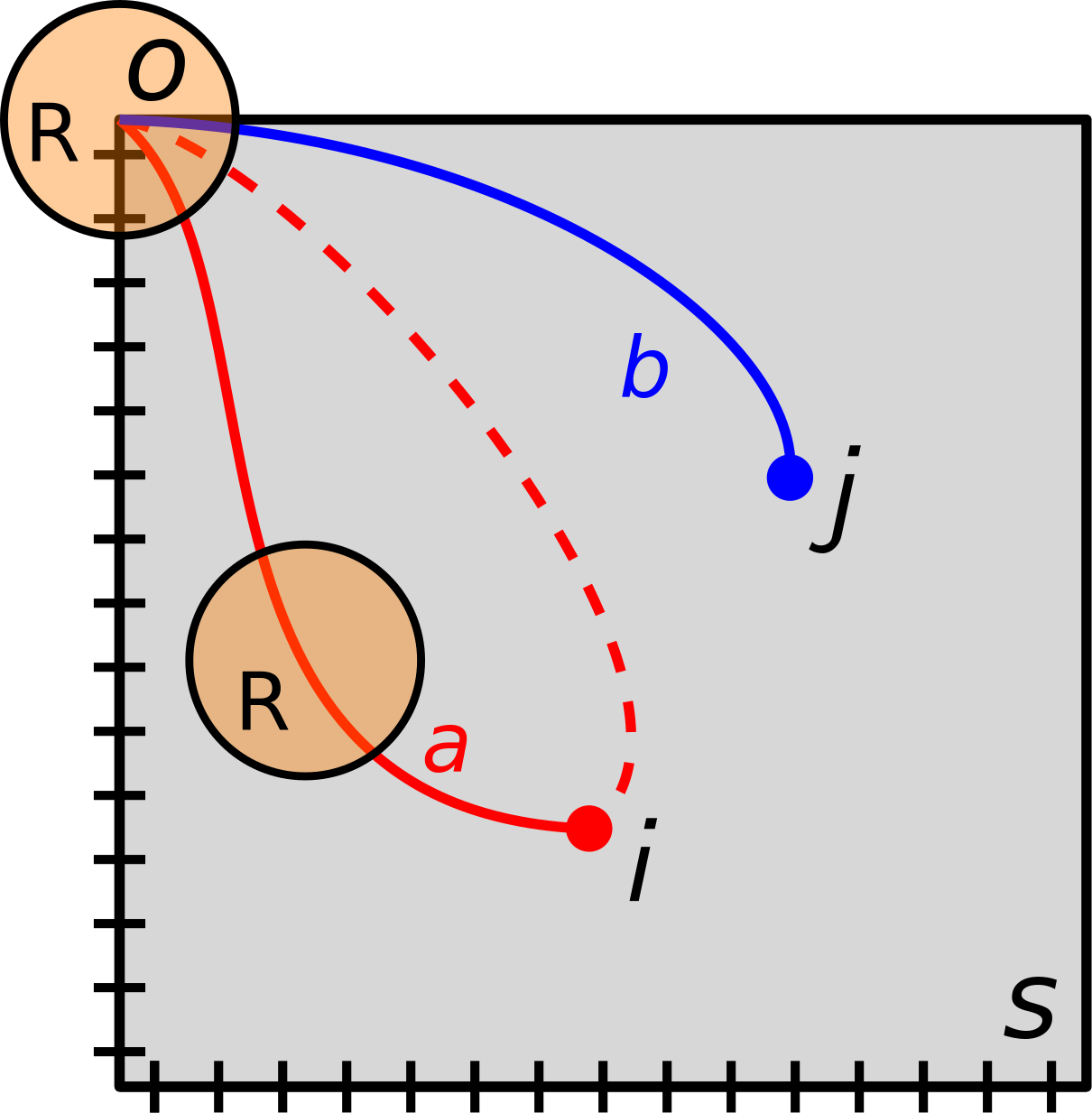}
				\caption{\label{fig:2ptstate} $|G;i^a j^b\rangle$ %
				}%
		\end{subfigure}\!\!\!
		\begin{subfigure}[t]{.59\linewidth}
			\includegraphics[width=\linewidth,left]{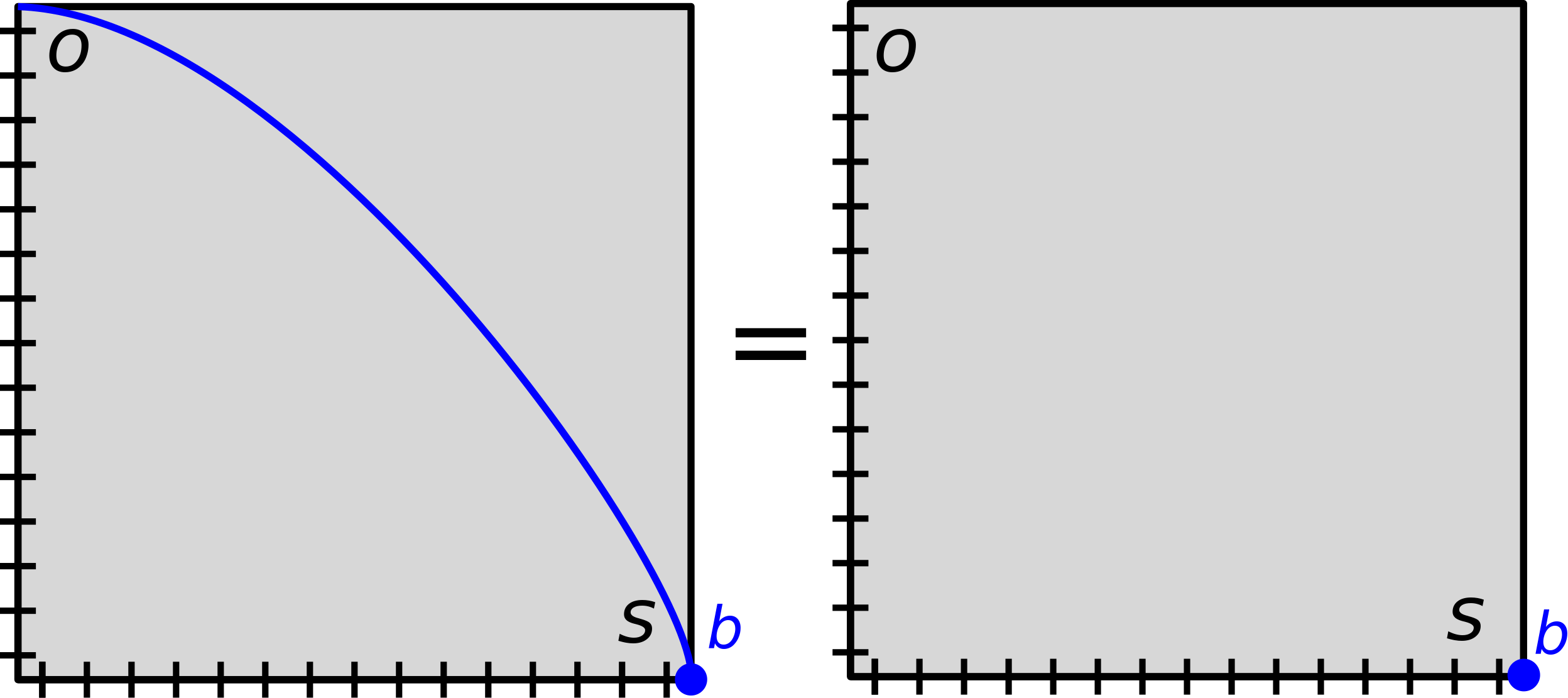}
		\caption{\label{fig:altcorner} $\hat{\psi}^+_{s,b}\ket{G}=\hat{y}^{+}_{s,b}\ket{G}=\hat{U}'_{s,b}\ket{G}$}
	\end{subfigure}
\caption{\label{fig:solvablemodel} Properties of the emergent paraparticles in the 2D solvable model. 
(a) A two-particle state %
$|G;i^a j^b\rangle\equiv \hat{\psi}^+_{i,a}\hat{\psi}^+_{j,b}\ket{G}$. 
A paraparticle at position $i$ with internal state $a$ is created by $\hat{\psi}^+_{i,a}$, a string operator connecting site $i$ to the corner $\oA$. Importantly, 
the action of $\hat{\psi}^+_{i,a}$ on $\ket{G}$ does not depend on the shape of the string, implying that
the Referees cannot detect the presence of such a string operator using any local measurement far from $i$~(e.g., any local measurements in the orange circles); %
(b) Paraparticles can also be locally created at $\oB$, using $\hat{U}^{\prime}_{\oB,b}$, since the string operator $\hat{W}_{bc}$ in $\hat{\psi}^+_{i,a}$ in Eq.~\eqref{eq:JWT_string_2D} connecting $\oA$ and $\oB$ acts trivially on $\ket{G}$. %
}
\end{figure}

		Within the zero vortex sector $\Phi_0$, the interacting quantum spin Hamiltonian $\hat{H}$ can be exactly mapped to a free paraparticle Hamiltonian,  %
		using the generalized Jordan-Wigner transformation introduced in Ref.~\cite{wang2023para}. As illustrated in Fig.~\ref{fig:solvablemodel}, the paraparticle creation operator  $\hat{\psi}^+_{i,a}$ is a matrix product operator~\cite{Bultinck2017AnyonMPOA} acting on a string connecting the site $i$ to the corner $\oA$. 
		For example, for the purple string $\Gamma$ in Fig.~\ref{fig:mainlattice}, $\hat{\psi}_{i,a}^+$  is defined as
		\begin{equation}\label{eq:JWT_string_2D}
			\hat{\psi}_{i,a}^+%
			=
			\begin{tikzpicture}[baseline={([yshift=.4ex]current bounding box.center)}, scale=.74]
				\Tpmatrix{0}{0}{\oA}\umatrix{1}{0}{v_R^+}{1}\Tpmatrix{2}{0}{2}
				\Tpmatrix{3}{0}{4}\umatrix{4}{0}{u_R^+}{5}\Tpmatrix{5}{0}{6}\umatrix{6}{0}{v_L^+}{7}%
				\ytriangle{7}{0}{+}{i}
				\node at (-0.5,0) [left=-.3mm] {\small $a$};
			\end{tikzpicture}.%
		\end{equation}
		Physically, $\hat{\psi}^+_{i,a}$ creates a paraparticle with internal state $a$ at position $i$. 
		These paraparticle operators satisfy several important properties which we summarize below. 
		
		First, we have 
		\begin{equation}\label{eq:psistring_commute_H}
			[\hat{\psi}_{i,a}^+,\hat{A}_\nu]=[\hat{\psi}_{i,a}^+,\hat{B}_p]=[\hat{\psi}_{i,a}^+,\hat{h}_{jk}]=0,
		\end{equation}
		for any $\nu,p$ and $i\notin\{j,k\}$, which can be shown straightforwardly using a tensor network derivation~\cite{wang2023para}. %
		Eq.~\eqref{eq:psistring_commute_H} implies that the action of $\hat{\psi}_{i,a}^+$ leaves the zero-vortex sector $\Phi_0$ invariant. 
		Moreover, the fact that the string operator in $\hat{\psi}_{i,a}^+$ commutes with all the local terms in $\hat{H}$ implies that $\hat{\psi}_{i,a}^+$ does not create any excitation %
		anywhere beyond the site $i$. %

		Although each paraparticle operator $\hat{\psi}^+_{i,a}$ is defined on a specific string, its action on the zero-vortex sector $\Phi_0$ does not depend on the geometric shape of the string, only on its endpoints, which is reminiscent of the familiar fact in the toric code model~\cite{kitaev2003fault} that the geometric shape of the anyon string operators can be smoothly deformed. As illustrated in Fig.~\ref{fig:solvablemodel}, this shape-independence property implies that the string operator in  $\hat{\psi}^+_{i,a}$  is not locally detectable. This fact is crucial for the winning strategy using paraparticles: if Alice and Bob hold paraparticles inside their circle areas, the Referee will not be able to detect any excitations using local measurements anywhere beyond the circle areas. 
		
		In the zero vortex sector, the operators $\{\hat{\psi}^\pm_{i,a}\}$ satisfy %
		the $R$-matrix commutation relations~\cite{wang2023para} for paraparticles
		\begin{eqnarray}\label{eq:fundamental_Rcommu}
			\hat{\psi}^-_{i,a} \hat{\psi}^+_{j,b}&=&\sum_{cd}R^{ac}_{bd} \hat{\psi}_{j, c}^+ \hat{\psi}_{i,d}^-+\delta_{ab}\delta_{ij},\nonumber\\%
			\hat{\psi}^+_{i,a} \hat{\psi}^+_{j,b}&=&\sum_{cd}R^{cd}_{ab} \hat{\psi}_{j,c}^+ \hat{\psi}_{i,d}^+,\nonumber\\%
			\hat{\psi}^-_{i,a} \hat{\psi}^-_{j,b}&=&\sum_{cd}R^{ba}_{dc} \hat{\psi}_{j,c}^- \hat{\psi}_{i,d}^-,%
		\end{eqnarray}
		where $\hat{\psi}^-_{i,a}=(\hat{\psi}^+_{i,a})^\dagger$~\footnote{The hermiticity condition $\hat{\psi}^-_{i,a}=(\hat{\psi}^+_{i,a})^\dagger$ only holds if the $R$-matrix is unitary~\cite{wang2023para}, which is true for the $R$-matrix considered in this paper. Indeed, all the $R$-matrices known to be realizable in $d>1$ quantum spin systems are unitary. Furthermore, due to the physical meaning of the $R$-matrix encoded in Eq.~\eqref{eq:exchangestatR-npt} and its role in the winning strategy, it is natural to assume that only unitary $R$-matrices are relevant for the challenge game in this paper.  }. 
		The %
		tensor $R^{ab}_{cd}=\!\!\begin{tikzpicture}[baseline={([yshift=-.6ex]current bounding box.center)}, scale=0.45]
			\Rmatrix{0}{0}{R}
			\node  at (-1.5*\AL,\AL) {\footnotesize $a$};
			\node  at (1.5*\AL,\AL) {\footnotesize $b$};
			\node  at (-1.5*\AL,-\AL) {\footnotesize $c$};
			\node  at (1.5*\AL,-\AL) {\footnotesize $d$};
		\end{tikzpicture}$   satisfies the Yang-Baxter equation%
		\begin{alignat}{3}\label{eq:YBE}
			\begin{tikzpicture}[baseline={([yshift=-.8ex]current bounding box.center)}, scale=0.5]
				\Rmatrix{0}{\AL}{R}
				\Rmatrix{0}{-\AL}{R}
				\node  at (-\AL,2.6*\AL) {\footnotesize $a$};
				\node  at (\AL,2.6*\AL) {\footnotesize $b$};
				\node  at (-\AL,-2.6*\AL) {\footnotesize $c$};
				\node  at (\AL,-2.6*\AL) {\footnotesize $d$};
			\end{tikzpicture}&=
			\begin{tikzpicture}[baseline={([yshift=-.8ex]current bounding box.center)}, scale=0.5]
				\draw[thick] (-\AL,-2*\AL) -- (-\AL,2*\AL);
				\draw[thick] (\AL,-2*\AL) -- (\AL,2*\AL);
				\node  at (-\AL,2.6*\AL) {\footnotesize $a$};
				\node  at (\AL,2.6*\AL) {\footnotesize $b$};
				\node  at (-\AL,-2.6*\AL) {\footnotesize $c$};
				\node  at (\AL,-2.6*\AL) {\footnotesize $d$};
				\node  at (-1.5*\AL,0*\AL) {\footnotesize $\delta$};
				\node  at (1.5*\AL,0*\AL) {\footnotesize $\delta$};
			\end{tikzpicture},&~~~
			\begin{tikzpicture}[baseline={([yshift=-.8ex]current bounding box.center)}, scale=0.5]
				\Rmatrix{-\AL}{2*\AL}{R}
				\Rmatrix{\AL}{0}{R}
				\Rmatrix{-\AL}{-2*\AL}{R}
				\draw[thick] (-2*\AL,-\AL) -- (-2*\AL,\AL);
				\draw[thick] (2*\AL,\AL) -- (2*\AL,3*\AL);
				\draw[thick] (2*\AL,-\AL) -- (2*\AL,-3*\AL);
				\node  at (-2*\AL,3.6*\AL) {\footnotesize $a $};
				\node  at (0*\AL,3.6*\AL) {\footnotesize $b$};
				\node  at (2*\AL,3.6*\AL) {\footnotesize $c$};
				\node  at (-2*\AL,-3.7*\AL) {\footnotesize $d$};
				\node  at (0*\AL,-3.7*\AL) {\footnotesize $e$};
				\node  at (2*\AL,-3.7*\AL) {\footnotesize $f$};
			\end{tikzpicture}
			&=\begin{tikzpicture}[baseline={([yshift=-.8ex]current bounding box.center)}, scale=0.5]
				\Rmatrix{\AL}{2*\AL}{R}
				\Rmatrix{-\AL}{0}{R}
				\Rmatrix{\AL}{-2*\AL}{R}
				\draw[thick] (2*\AL,-\AL) -- (2*\AL,\AL);
				\draw[thick] (-2*\AL,\AL) -- (-2*\AL,3*\AL);
				\draw[thick] (-2*\AL,-\AL) -- (-2*\AL,-3*\AL);
				\node  at (-2*\AL,3.6*\AL) {\footnotesize $a$};
				\node  at (0*\AL,3.6*\AL) {\footnotesize $b$};
				\node  at (2*\AL,3.6*\AL) {\footnotesize $c$};
				\node  at (-2*\AL,-3.7*\AL) {\footnotesize $d$};
				\node  at (0*\AL,-3.7*\AL) {\footnotesize $e$};
				\node  at (2*\AL,-3.7*\AL) {\footnotesize $f$};
			\end{tikzpicture},\\
			R^2&=\mathds{1},&\quad R_{12}R_{23}R_{12}&=R_{23}R_{12}R_{23}.\nonumber
		\end{alignat}
		and therefore defines a representation of the symmetric group $S_n$. We will see later that $R^{cd}_{ab}$ 
		defines the exchange statistics of the paraparticles created by  $\hat{\psi}^+_{i,a}$ in the sense of Eq.~\eqref{eq:exchangestatR-npt},  %
		which follows from Eq.~\eqref{eq:fundamental_Rcommu}.

		Finally, in the zero vortex sector $\Phi_0$, $\hat{H}_2$ can be rewritten as a free paraparticle Hamiltonian
		\begin{equation}\label{eq:freeparaH2}
			\hat{H}_2 =-\sum_{\langle ij\rangle, 1\leq a\leq m}  (J_{ij}\hat{\psi}^+_{j,a}\hat{\psi}^-_{i,a}+\mathrm{h.c.})-\sum_{l} \mu_l \hat{n}_l.
		\end{equation}
		Note that since $\hat{H}_1$ is just a constant in $\Phi_0$, we have $\hat{H}=\hat{H}_2+\mathrm{const.}$, i.e., $\hat{H}$ is mapped to free paraparticles in $\Phi_0$.  
		In this paper we assume that the free parameters $\{J_{ij},\mu_l\}$ are chosen such that the paraparticles have a topologically trivial band structure with a nonzero band gap~\footnote{
			Actually, we expect that any gapped phase of this model can be used to pass the challenge, since emergent paraparticles exist regardless of the parameters $\{J_{ij},\mu_l\}$~(the exact mapping to free paraparticles presented in this section does not depend on these parameters). However, if the paraparticle spectrum has a topologically nontrivial band structure with Chern number $\nu\neq 0$, then the system will be in a chiral topological phase. This can be  interesting to study in its own right, but for the purpose of winning the challenge, it causes two complications. First, chiral topological phases of this model are not frustration free, therefore one can only pass a generalized version of the game in which the frustration-free requirement is relaxed, as we mentioned in Sec.~\ref{sec:relax_FF}. Second, for chiral topological phases, one cannot use the hybrid open boundary condition as shown in Fig.~\ref{fig:mainlattice},  since gapless chiral edge modes violate the game requirement, so one need to choose $\oA$ and $\oB$ to be the positions of other types of point-like defects in the system.   
		}, 
		with the ground state $\ket{G}$ being the paraparticle vacuum, satisfying $\hat{\psi}^-_{i,a}\ket{G}=0$ everywhere. This happens, for example,  when
		\begin{equation}\label{eq:nu=0condition}
			-\mu_l=-\mu\gg J= |J_{ij}|, \text{ for all } i,j,l. 
		\end{equation}
		This guarantees that the model satisfies all the requirements of the game, as $\ket{G}$ is unique, gapped, and frustration-free, all of which can be rigorously proved~\cite{wang2023para}. The Referees can monitor the local ground state condition simply by measuring the local conserved quantities along with the paraparticle number operators $\hat{n}_l\equiv\sum_{a}  \hat{\psi}^+_{l,a}\hat{\psi}^-_{l,a}=\sum_{a}  \hat{y}^+_{l,a}\hat{y}^-_{l,a}$.%
		
		\subsection{Remarks on the open boundary condition}\label{sec:2DESM_OBC}
		As we emphasized in the main text, to win the challenge, the two points $\oA$ and $\oB$ have to be specially chosen such that a single paraparticle can be locally created and measured at both positions~[see Eqs.~\eqref{eq:localcreationatcorner} and \eqref{eq:localmeasurementatcorner}]. 
		With the hybrid boundary condition as shown in  Fig.~\ref{fig:mainlattice}, %
		we can choose $\oA$ and $\oB$ to be the  intersection points between the two different types of gapped boundaries of this model. 
		It is well-known that a topological phase can have several possible gapped boundaries with distinct topological properties~\cite{Kitaev2012gappedboundary}. For example, the toric code model has two possible gapped boundaries~\cite{bravyi1998TCBoundary}, one is called a smooth boundary, where one can fuse a magnetic flux into vacuum, and a rough boundary, where one can fuse an electric charge into vacuum. These two types of gapped boundaries can be defined for more general Kitaev's quantum
		double model based on any finite dimensional $\C^*$-Hopf algebra. This fact also applies to our model, which is an extension of a Kitaev's quantum double model based on a special 64-dimensional Hopf algebra~\cite{wang2023para}. As shown in Fig.~\ref{fig:mainlattice}, the western and southern boundaries are rough, where the $\hat{B}_p$ terms are truncated, while the eastern and northern boundaries are smooth where the $\hat{A}_\nu$ terms are truncated. If we use such a hybrid gapped boundary in the toric code model, then one can fuse a fermion into vacuum at the corners $\oA$ and $\oB$, which are the intersection points between the smooth and rough boundaries. Analogously, in our model, this hybrid boundary condition allows one to fuse paraparticles into vacuum at the corners $\oA$ and $\oB$, which is crucial for winning the challenge. 
		
		More specifically, it is clear from Eq.~\eqref{eq:JWT_string_2D} that $\hat{\psi}^+_{\oA,a}=\hat{y}^+_{\oA,a}$, %
		meaning that one can use the local operator $\hat{y}^+_{\oA,a}$ to create a single paraparticle with internal state $a$ at the corner $\oA$. %
		By symmetry, it is natural to expect that paraparticles can also be locally created at the corner $\oB$~\footnote{At first glance, it may appear that the paraparticle operators $\hat{\psi}^+_{i,a}$ in Eq.~\eqref{eq:JWT_string_2D} distinguish between the two corners $\oA$ and $\oB$, as their MPO strings start from $\oA$ instead of $\oB$. However, this is simply a matter of convention. In this model there is no essential difference between the two corners $\oA$ and $\oB$, and we can alternatively define the MPO JWT to start from $\oB$ as well.}. 
		A more careful analysis is given as follows. As illustrated in Fig.~\ref{fig:altcorner}, the paraparticle creation operator $\hat{\psi}^+_{\oB,b}$ is a non-local string operator connecting $\oA$ to $\oB$, and from  Eq.~\eqref{eq:JWT_string_2D}, we have $\hat{\psi}^+_{\oB,b}=\sum_{c}\hat{W}_{bc}\hat{y}^{+}_{\oB,c}$, where $\hat{W}_{bc}$ is the JW string connecting $\oA$ to $\oB$. 
		Ref.~\cite{wang2023para} shows that $\hat{W}_{bc}\ket{G}=\delta_{bc}\ket{G}$, %
		leading to 
		\begin{equation}\label{eq:psioByoB}
			\hat{\psi}^+_{\oB,b}\ket{G}=\sum_{c}\hat{y}^{+}_{\oB,c}\hat{W}_{bc}\ket{G}=\hat{y}^{+}_{\oB,b}\ket{G}.
		\end{equation}
		Therefore, one can use the local operator $\hat{y}^{+}_{\oB,b}$ to create a paraparticle at the corner $\oB$. 
		
		\subsection{Universal properties of the emergent paraparticles}\label{sec:2DESM-universal}
		The above second quantization formulation directly leads to the simple description of emergent paraparticles presented in Eqs.~(\ref{eq:TPdegenerate}-\ref{eq:localmeasurementatcorner}), and we give their relation below:\\ %
		(1) %
		The basis states for the $n$-particle subspace $\Hil_{i_1i_2 \ldots i_n}$ are defined as %
		\begin{equation}\label{eq:basis_n_particle_space}
			\ket{G;i_1^{a_1} i_2^{a_2} \ldots i_n^{a_n}}=\hat{\psi}^+_{i_1,a_1}\hat{\psi}^+_{i_2,a_2}\ldots\hat{\psi}^+_{i_n,a_n}\ket{G}.
		\end{equation}
		(2) Topological degeneracy of $\Hil_{i_1i_2 \ldots i_n}$ directly follows from the fact that the paraparticle created by the string operator $\hat{\psi}^+_{i,a}$ is a topological excitation with quantum dimension $m$, and Eq.~\eqref{eq:TPdegenerate} holds for topologically degenerate states in general. \\
		(3) %
		Paraparticles can be moved using local unitaries of the form $\hat{U}_{ij}=e^{i\Delta t (\hat{e}_{ij}+ \hat{e}_{ji})}$, where 
		$\hat{e}_{ij}\equiv \sum^m_{a=1} \hat{\psi}^+_{i,a}\hat{\psi}^-_{j,a}$
		are paraparticle tunneling operators~[e.g., when $i,j$ are neighbors, we have $\hat{e}_{ij}=\hat{h}_{ij}$, the three-body term defined in Eq.~\eqref{def:3bodyterm}]. %
		Eq.~\eqref{eq:paraparticlemove} follows from 
		\begin{equation}\label{eq:Uij_action}
			\hat{U}_{ij}\hat{\psi}^+_{i,a}\hat{U}_{ij}^\dagger=\hat{\psi}^+_{j,a},\quad	\hat{U}_{ij}\hat{\psi}^+_{j,a}\hat{U}_{ij}^\dagger=\hat{\psi}^+_{i,a},\quad \forall a,
		\end{equation}
		which can be derived using Eq.~\eqref{eq:fundamental_Rcommu}. \\
		(4) Exchange statistics in Eq.~\eqref{eq:exchangestatR} %
		follows from the definition in Eq.~\eqref{eq:basis_n_particle_space} and the second line of Eq.~\eqref{eq:fundamental_Rcommu}. In the $n$-particle subspace $\Hil_{i_1i_2 \ldots i_n}$, Eq.~\eqref{eq:exchangestatR} generalizes to  
		\begin{equation}\label{eq:exchangestatR-npt}
			\ket{G;\ldots i_k^{a}i_{k+1}^{b}\ldots }=\sum_{a',b'}R^{b'a'}_{ab}\ket{G;\ldots i_{k+1}^{b'}i_k^{a'} \ldots},
		\end{equation}
		for $k=1,\ldots,n-1$, where $\ldots$ collectively denotes other labels that are unaffected by the exchange. 
		Eq.~\eqref{eq:exchangestatR-npt} completely determines the basis transformation between $\Hil_{i_1 i_2 \ldots i_n}$ and $\Hil_{j_1 j_2 \ldots j_n}$, where  $j_1,\ldots,j_n$ is a permutation of $i_1,\ldots,i_n$, 
		since any permutation can be written as a composition of a finite sequence of neighboring swaps of the form in Eq.~\eqref{eq:exchangestatR-npt}. \\
		(5) %
		The local unitary operators $\hat{U}_{\oA,a}$, $\hat{U}'_{\oB,a}$ in Eq.~\eqref{eq:localcreationatcorner} can be constructed as %
		\begin{eqnarray}\label{eq:localcreation_Udef}
			\hat{U}_{\oA,a}&=&e^{i\Delta t (\hat{y}^+_{\oA,a}+\hat{y}^-_{\oA,a})},\nonumber\\
			\hat{U}'_{\oB,a}&=&e^{i\Delta t (\hat{y}^{+}_{\oB,a}+\hat{y}^{-}_{\oB,a})},
		\end{eqnarray}
		for a suitable $\Delta t$ such that $\hat{U}_{\oA,a}\ket{G}=\hat{y}^+_{\oA,a}\ket{G}=\hat{\psi}^+_{\oA,a}\ket{G}$. Note that both are unitary and act locally on the sites at $\oA, \oB$, respectively.  \\
		(6) %
		The local observables in Eq.~\eqref{eq:localmeasurementatcorner} can be constructed as 
		\begin{eqnarray}\label{eq:def_color_operator}
			\hat{O}_\oA&=&\sum^4_{a=1} a~\hat{y}^+_{\oA,a}\hat{y}^-_{\oA,a},\nonumber\\
			\hat{O}'_\oB&=&\sum^4_{a=1} a~\hat{y}^{+}_{\oB,a}\hat{y}^{-}_{\oB,a}.
		\end{eqnarray}
		Eq.~\eqref{eq:localmeasurementatcorner} can be proved using Eq.~\eqref{eq:fundamental_Rcommu} and the fact that $\hat{O}_\oA$, $\hat{O}'_\oB$ act locally on the corners $\oA$, $\oB$, respectively~\cite{wang2023para}. For example, the first line can be proved by noticing that
		\begin{eqnarray}\label{eq:localmeasurementatcorner_proof1st}
			\hat{\psi}_{i_1,a}^-\ket{G;i_1^{a_1} \ldots i_n^{a_n}}&=&\hat{\psi}_{i_1,a}^-\hat{\psi}^+_{i_1,a_1}\ldots\hat{\psi}^+_{i_n,a_n}\ket{G}\\
			&=&\delta_{a,a_1}\hat{\psi}^+_{i_2,a_2}\ldots\hat{\psi}^+_{i_n,a_n}\ket{G},\nonumber
		\end{eqnarray}
		where we use the first line of Eq.~\eqref{eq:fundamental_Rcommu}  and the assumption that $i_1,i_2,\ldots,i_n$ are mutually different. Taking $i_1=\oA$ in Eq.~\eqref{eq:localmeasurementatcorner_proof1st} and using the fact that $\hat{\psi}^\pm_{\oA,a}=\hat{y}^\pm_{\oA,a}$, we obtain the first line of Eq.~\eqref{eq:localmeasurementatcorner}. The second line of Eq.~\eqref{eq:localmeasurementatcorner} is proved as follows: for $i_n=\oB$, we have
		\begin{eqnarray}\label{eq:localmeasurementatcorner_proof2st}
			\hat{O}'_{\oB}\ket{G;i_1^{a_1} \ldots i_n^{a_n}}&=&\hat{O}'_{\oB}\hat{\psi}^+_{i_1,a_1}\ldots\hat{\psi}^+_{i_n,a_n}\ket{G}\nonumber\\
			&=&\hat{\psi}^+_{i_1,a_1}\ldots \hat{O}'_{\oB}\hat{\psi}^+_{i_n,a_n}\ket{G}\nonumber\\
			&=&\hat{\psi}^+_{i_1,a_1}\ldots \hat{O}'_{\oB}\hat{y}^+_{\oB,a_n}\ket{G}\nonumber\\
			&=&a_n\ket{G;i_1^{a_1} \ldots i_n^{a_n}}, 
		\end{eqnarray}
		where in the second  line we used the assumption that $i_1,\ldots,i_{n-1}$ are different from $i_n=\oB$,  in the third line we used Eq.~\eqref{eq:psioByoB}, and in the last line we used  
		\begin{equation}\label{eq:OobAction1pt}
			\hat{O}'_{\oB}\hat{y}^+_{\oB,b}\ket{G}=b\hat{y}^+_{\oB,b}\ket{G}
		\end{equation} 
		which follows from the algebra of $\{\hat{y}^\pm_{\oB,a}\}_{a=1}^m$~\footnote{Indeed, at \textit{any single site} $i$, the local operators $\{\hat{y}^\pm_{i,a}\}_{a=1}^m$ satisfy the same commutation relations~[Eq.~\eqref{eq:fundamental_Rcommu}] as the nonlocal paraparicle operators $\{\hat{\psi}^\pm_{i,a}\}_{a=1}^m$. Therefore, Eq.~\eqref{eq:OobAction1pt} follows from the $n=1$ case of Eq.~\eqref{eq:localmeasurementatcorner_proof1st}.}.

\subsection{Categorical description of the emergent paraparticles and the solvable models}\label{sec:cat_description}
Tensor category theory~\cite{TenCat_EGNO} has been widely adopted in the literature to describe the universal properties of topological excitations in topological phases of matter~\cite{kitaev2006anyons,kong2014braided}. 
In this section we briefly mention the categorical description of paraparticles in this 2D exactly solvable model.
In the accompanying Mathematica code~\cite{MmaCode}, we give three different ways to construct the SFC $\mathcal{C}$ describing paraparticles in this model. 
First, we can construct $\mathcal{C}$ as $\mathrm{Rep}[\Hil_{64}]$, the category of finite dimensional representations of the 64-dimensional triangular Hopf algebra $\Hil_{64}$ used to construct the 2D solvable model~\cite{wang2023para}. Meanwhile, we also have 
$\mathcal{C}\cong\mathrm{sRep}(G,z)$, where $G=D_4\ltimes Z_2^{\times 3}$ is a group of order 64 and $z\in G$ is a central element.  

More generally, in the same Mathematica package~\cite{MmaCode} we provide an algorithm to reconstruct the underlying SFC $\mathcal{C}$ from a given $R$-matrix satisfying certain constraints~\footnote{Not every type of parastatistics can be described by an SFC--the $R$-matrix has to satisfy certain condition for this to be possible.  There are several equivalent ways to formulate this condition, which will be presented in a future work. Here we mention a simple tensor network description of this condition: when regarded as a two-qudit quantum gate, $R$ generates a Hopf algebra solvable circuit introduced in Ref.~\cite{wang2024hopf}. More precisely, there exists four-index tensors $\rho,v$ such that the gate $U=R$ satisfies Eq.~(3) in Ref.~\cite{wang2024hopf} for any $n\geq 1$. As an example, one can use this condition to quickly rule out the $R$-matrix $R=-\mathds{1}_{m^2}$ in Ex.~3 of Tab.~I of Ref.~\cite{wang2023para}--indeed, this $R$-matrix cannot be produced by any SFC and is unlikely to be realizable in any gapped phase of matter in dimension $d>1$. If one inputs such an $R$-matrix into the algorithm in the accompanying code~\cite{MmaCode}, it will run forever.}. 
For example, in the accompanying code~\cite{MmaCode}, we apply this algorithm to  the $R$-matrix in Eq.~\eqref{eqApp:seth-R}, and obtain an SFC $\mathcal{C}$ which turns out to be equivalent to $\mathrm{Rep}[\Hil_{64}]$ discussed above.
We emphasize that the SFC $\mathcal{C}$ constructed by this algorithm describe the universal properties of the $R$-paraparticles independent of its microscopic realization~(regardless of whether it is realized in 2D or 3D). A caveat, however, is that $\mathcal{C}$ constructed by this algorithm only describes the subcategory generated by $R$-paraparticles in the hosting topological phase, since the latter generally have other type of particles not contained in $\mathcal{C}$
~\footnote{
In particular, the point-like excitation/defect $\sigma$ mentioned in  Eqs.~\eqref{eq:SFCdescriptiongame} and \eqref{eq:sigmapsifusion} of the main text is often not contained in the SFC $\mathcal{C}$ constructed by this algorithm. 
For the $R$-paraparticle described by Eq.~\eqref{eqApp:seth-R} of the main text~(with the ``$+$'' sign), $\sigma$ can be found in a larger SFC
$\mathcal{C}'=\mathrm{Rep}(G')$ that includes $\mathcal{C}$ as a full subcategory, where $G'$ is a $Z_2$-central extension of $G=D_4\ltimes Z_2^{\times 3}$, given in Eq.~\eqref{def:G128}~(with $c$ being the central element of order $2$), and $\sigma\in \mathcal{C}'$ is a 8-dimensional projective representation of $G$ that is lifted to a linear representation of $G'$. It is straightforward to show that $\mathcal{C}'$ has a fusion rule of the form in Eq.~\eqref{eq:sigmapsifusion}, and therefore can be used to win the 3D game without any boundary, as shown in Sec.~\ref{SI:3Dmodelparaparticle}. 
}. 

One may also be interested in the categorical description of the 2D exactly solvable model presented in Sec.~\ref{sec:2DESMreview_Hamiltonian}. This 2D model not only hosts emergent paraparticles, but also non-Abelian anyons, %
which are described by unitary modular tensor categories~(UMTC).
The topological order of the 2D solvable model depends on the Chern  number $\nu$ of the paraparticle spectrum, which in turn depends on the coupling constants $J_{ij},\mu_i$ of the model, and the complete classification of phases in this model is still unknown up to now. However, we know that in any gapped phase of this model, the SFC  %
$\mathcal{C}=\mathrm{sRep}(G,z)$ 
is always included as a full subcategory of the UMTC describing the whole system, which is just a categorical way of saying that $R$-paraparticles emerge in any gapped phase of this model~(indeed, they emerge in gapless phases of this model as well, which  are not described by UMTCs). Furthermore, we known that when $\nu=0$~[e.g., when Eq.~\eqref{eq:nu=0condition} is satisfied]%
this model is in the same topological phase as Kitaev's quantum double model~\cite{kitaev2003fault} based on the group $G$, described by the UMTC $\mathrm{Rep}[D(G)]\cong Z[\Rep(G)]$, the representation category of the Drinfeld double $D(G)$, or, equivalently, the Drinfeld center of $\Rep(G)$. 

\section{A 3D solvable model with emergent paraparticles}\label{SI:3Dmodelparaparticle}
In this section we briefly describe a 3D exactly solvable model hosting emergent $R$-paraparticles. This model 
is a 3D deconfined lattice gauge theory based on a specially chosen finite group $\Gamma$, and it realizes an example of the special class of SFCs described in the main text, i.e., the system hosts topological quasiparticles $\sigma,\bar{\sigma},\psi$ with fusion rules of the form in Eq.~\eqref{eq:sigmapsifusion}, such that the $R$-matrix defined by Eq.~\eqref{eq:SFCdescriptiongame} is non-trivial. In the following we first give a quick review of general 3D deconfined lattice gauge theories~\cite{FradkinShenker1979,kogutIntroductionLatticeGauge1979, wangNonAbelianStringParticle2015,Zhu3DTPOModels,LanKongWen3DAB,LanWen3DEF} in Sec.~\ref{SI:GTreview}, and then in Sec.~\ref{SI:RparaGT} we derive emergent $R$-parastatistics in a specific model with a specially designed gauge group $\Gamma$. 

\subsection{A quick review of 3D deconfined solvable lattice gauge theory}\label{SI:GTreview}
\begin{figure}
	\centering\includegraphics[width=.9\linewidth]{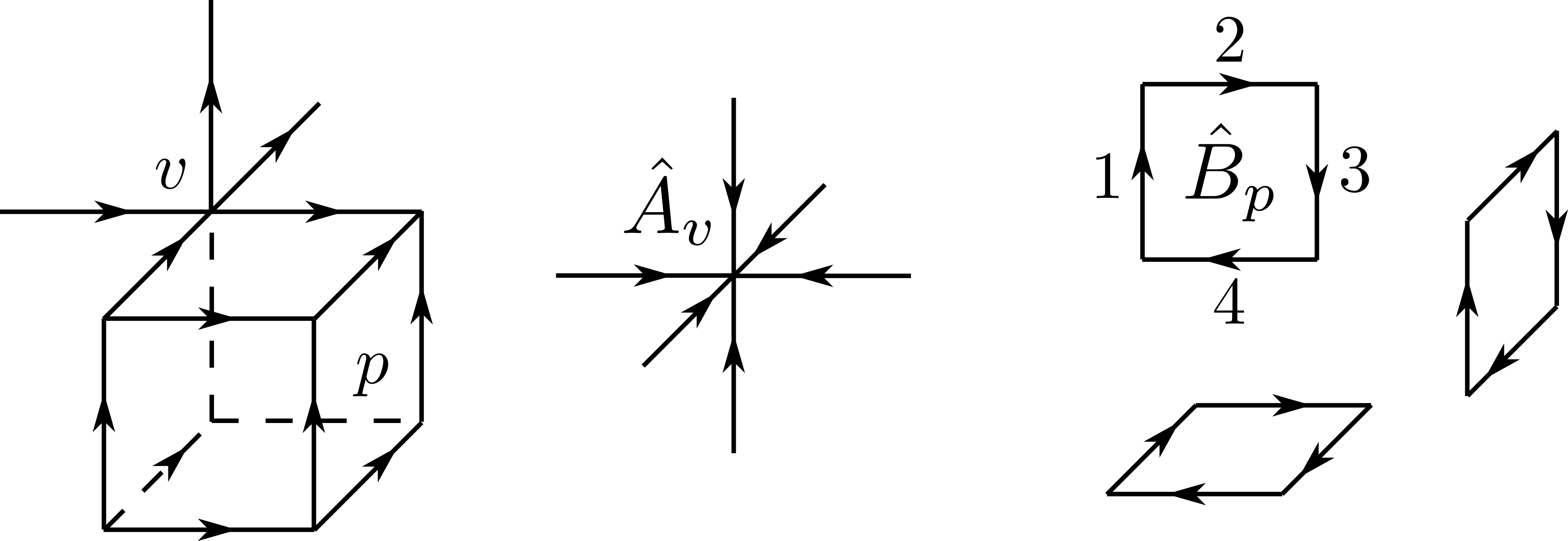} 
	\caption{\label{fig:3DGaugeLattice} Definition of the vertex term $\hat{A}_v$ and the plaquette term $\hat{B}_p$ in Eq.~\eqref{eq:3DLGT}. 
	}
\end{figure}

We define the system on a 3D oriented cubic lattice, where a qudit sits on each link, as shown in Fig.~\ref{fig:3DGaugeLattice}. 
We label the basis states of each qudit by elements of a finite group $\Gamma$~(directly generalizing Kitaev's quantum double models~\cite{kitaev2003fault} to 3D). The lattice is oriented as shown in Fig.~\ref{fig:3DGaugeLattice}, and reversing the direction of a link corresponds to applying $g\to g^{-1}$. 
The Hamiltonian is 
\begin{eqnarray}\label{eq:3DLGT}
\hat{H}_0&=&-\sum_{v}\hat{A}_v-\sum_{p}\hat{B}_p,\nonumber\\	
\hat{A}_v&=&\frac{1}{|\Gamma|}\sum_{g\in \Gamma}\hat{L}_v^g,\quad \hat{L}_v^g=\prod_{j\in v}\hat{L}_j^g,\nonumber\\
\hat{B}_p&=&\sum_{g_1 g_2 g_3 g_4=1}\hat{\delta}_1^{g_1}\hat{\delta}_2^{g_2}\hat{\delta}_3^{g_3}\hat{\delta}_4^{g_4},
\end{eqnarray}
where $v$ runs over all vertices $v$, and $p$ runs over all plaquettes $p$, and $\hat{A}_v$ and $\hat{B}_p$ are defined with respect to the orientation shown in Fig.~\ref{fig:3DGaugeLattice}. Here the single site operators $\hat{L}^g,\hat{\delta}^{g}$ are defined as
\begin{eqnarray}
\hat{L}^g\ket{h}&=&\ket{gh},\nonumber\\
\hat{\delta}^{g}\ket{h}&=&\delta_{g,h}\ket{h},
\end{eqnarray}
for any $h\in \Gamma$. It is straightforward to check that all the local terms $\hat{A}_v$ and $\hat{B}_p$ in  $\hat{H}_0$ are mutually commuting projection operators, and $\hat{H}_0$ has a unique, gapped, and frustration-free ground state  $\ket{G_0}$ on a topologically trivial manifold, satisfying
\begin{equation}
\hat{A}_v\ket{G_0}=\hat{B}_p\ket{G_0}=\ket{G_0},
\end{equation}
for all $v$ and $p$. 

All point-like quasiparticles of $\hat{H}_0$ are created by Wilson line operators.
Each simple Wilson line operator is labeled by an irrep $\psi$ of the gauge group $\Gamma$, %
represented as an MPO generated by  a 4-index tensor~\cite{kitaev2003fault,schuchPEPSGroundStates2010,Bultinck2017AnyonMPOA}
\begin{equation}\label{eq:PsiTensor}
	\adjincludegraphics[height=9ex,valign=c]{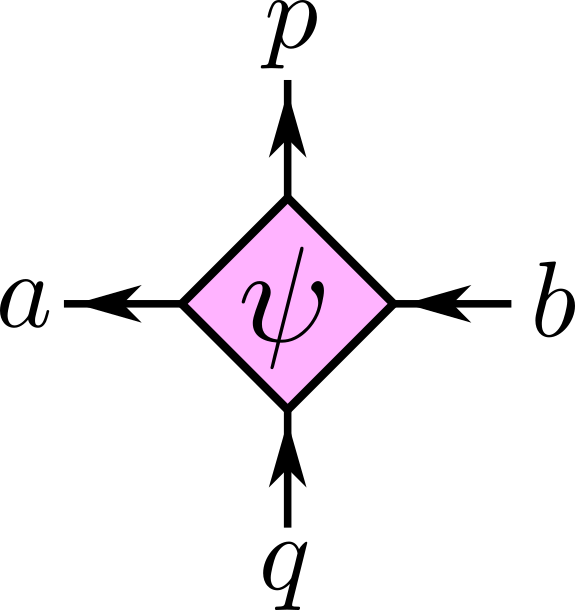}=\sum_{g\in\Gamma }[\psi(g)]_{ab} [\hat{\delta}^g]_{pq},
\end{equation}
where $p,q\in \Gamma$ labels the qudit basis, and $a,b\in\{1,2,\ldots,d_\psi\}$.  
For example, the Wilson line operator acting on four consecutive links is 
\begin{eqnarray}\label{eq:PsiLine}
	\adjincludegraphics[width=11ex,valign=c]{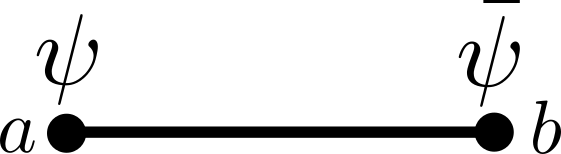}
	=\adjincludegraphics[height=6ex,valign=c]{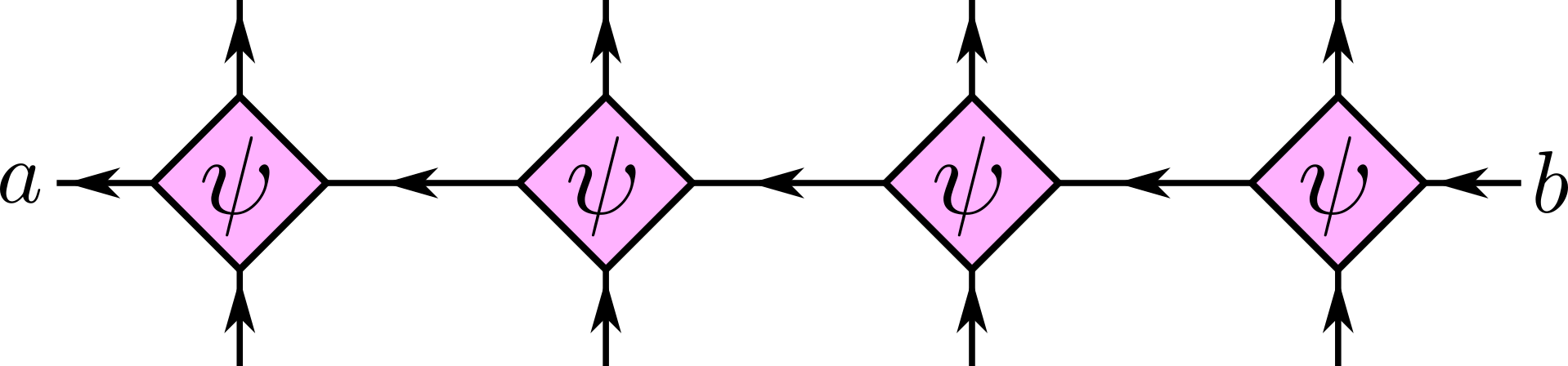}.%
\end{eqnarray}
When acting on the ground state $\ket{G_0}$, this Wilson line operator creates two deconfined quasiparticles at its two endpoints, one is $\psi$, and the other is $\bar{\psi}$. Let $\ket{\Psi^{(a,b)}_{v_1,v_2}}=\adjincludegraphics[width=11ex,valign=c]{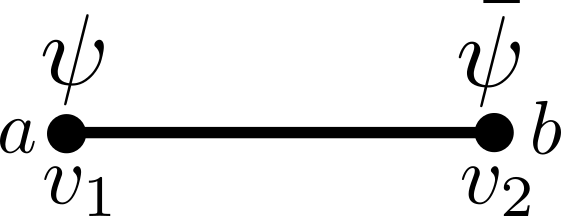}\ket{G_0}$. Then we have
\begin{eqnarray}\label{eq:localdofPsi}
\hat{L}^g_{v_1}\ket{\Psi^{(a,b)}_{v_1,v_2}}&=&\sum_{c}[\psi(g^{-1})]_{ac}\ket{\Psi^{(c,b)}_{v_1,v_2}},\nonumber\\
\hat{L}^g_{v_2}\ket{\Psi^{(a,b)}_{v_1,v_2}}&=&\sum_{c}[\psi(g)]_{cb}\ket{\Psi^{(a,c)}_{v_1,v_2}},
\end{eqnarray}
and it is straightforward to check that
\begin{eqnarray}
\hat{A}_v\ket{\Psi^{(a,b)}_{v_1,v_2}}&=&(1-\delta_{v,v_1}-\delta_{v,v_1})\ket{\Psi^{(a,b)}_{v_1,v_2}},\nonumber\\
\hat{B}_p\ket{\Psi^{(a,b)}_{v_1,v_2}}&=&\ket{\Psi^{(a,b)}_{v_1,v_2}},
\end{eqnarray}
i.e., $\ket{\Psi^{(a,b)}_{v_1,v_2}}$ only contains point-like excitations at vertices $v_1$ and $v_2$. 
Furthermore, the action of the Wilson line operator on $\ket{G_0}$ can be smoothly deformed:
\begin{equation}\label{eq:PsiLinedeform}
	\adjincludegraphics[width=11ex,valign=c]{Figures/ParaExchange/PsiLine2IndPos.png}\ket{G_0}
	=\adjincludegraphics[width=11ex,valign=c]{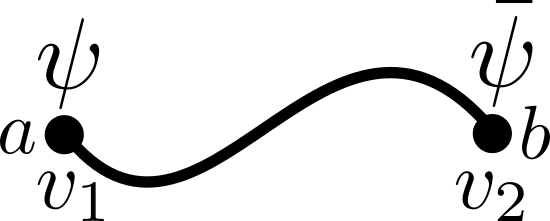}\ket{G_0},
\end{equation}
and is therefore invisible except at its two endpoints $v_1$ and $v_2$. For later convenience, we also define a local trapping potential for each quasiparticle $\psi$. Let 
\begin{equation}
\hat{A}^{(\psi,a)}_v=\sum_{g\in \Gamma}[\psi(g)]_{a,a}\hat{L}_v^g.
\end{equation}
Then it is straightforward to check that 
\begin{eqnarray}\label{eq:trappingpotential}
\hat{A}^{(\varphi,c)}_{v_1}\ket{\Psi^{(a,b)}_{v_1,v_2}}&=&C_1\delta_{ca}\delta_{\varphi\psi}\ket{\Psi^{(a,b)}_{v_1,v_2}},\nonumber\\
\hat{A}^{(\varphi,c)}_{v_2}\ket{\Psi^{(a,b)}_{v_1,v_2}}&=&C_2\delta_{cb}\delta_{\varphi\bar{\psi}}\ket{\Psi^{(a,b)}_{v_1,v_2}},
\end{eqnarray}
where $C_1,C_2$ are nonzero constants. This allows one to define a local trapping potential $\hat{V}^{(\psi,a)}_v$ for each  $\psi$, such that $\ket{\Psi^{(a,b)}_{v_1,v_2}}$ is the unique, gapped ground state of $\hat{H}_0+\hat{V}^{(\psi,a)}_{v_1}+\hat{V}^{(\bar{\psi},b)}_{v_2}$. 
Eq.~\eqref{eq:localdofPsi} implies that the labels  $a,b$ in $\ket{\Psi^{(a,b)}_{v_1,v_2}}$ correspond to local degrees of freedom that can be changed by local operations, and in this paper we do not care about such local degrees of freedom, so in the following we will forget about them and simply use $\adjincludegraphics[width=11ex,valign=c]{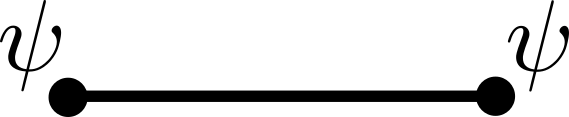}$ to mean
 $\adjincludegraphics[width=11ex,valign=c]{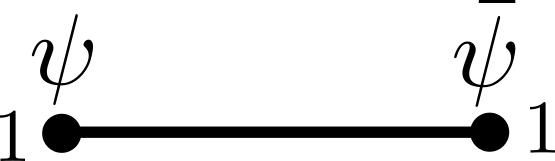}$, and similarly use $\hat{V}^{\psi}_v$ to mean $\hat{V}^{(\psi,1)}_v$.

\subsection{The special gauge group $\Gamma$ and emergent $R$-paraparticles}\label{SI:RparaGT}
We now choose a special gauge group $\Gamma$ such that the model defined in Eq.~\eqref{eq:3DLGT} hosts emergent $R$-paraparticles.   
The group $\Gamma$ has order $|\Gamma|=128$, and is generated by 
$r, s, z_1,z_2,z_3,z_4$ along with a central element $c$, subject to the relations
\begin{eqnarray}\label{def:G128}
	&& r^4=s^2=c^2=1,~ s r s^{-1}=r^{-1},\nonumber\\
	&& z_j^2=1,~	r z_j r^{-1}=z_{j+1},~ s z_j s^{-1}=c z_{\bar{j}}, \nonumber\\
	&& z_i z_j=c^{i-j} z_j z_i,~\text{ for }1\leq i,j\leq 4,
\end{eqnarray}
where $(\bar{1},\bar{2},\bar{3},\bar{4})=(2,1,4,3)$, and the subscript $j$ is understood modulo 4. More details on this group can be found in Ref.~\cite{wang2025secret}. This group 
$\Gamma$ is specially constructed such that there are $\sigma,\psi\in \Rep(\Gamma)$ with a fusion rule
\begin{equation}\label{eq:sigmapsifusion-0}
\sigma \times \psi =d_\psi~\sigma.
\end{equation}
This fusion rule implies an isomorphism between the corresponding irreps of $\Gamma$,
meaning that there exists a unitary matrix $V$ satisfying
\begin{equation}\label{eq:RepInterwiner}
	[\sigma(g) \otimes \psi(g)] \cdot V=V\cdot[I_m \otimes \sigma(g)],\quad \forall ~g\in \Gamma, 
\end{equation} 
where $m=d_\psi$, and $I_{m}$ is the $m\times m$ identity matrix, so that $I_m \otimes \sigma(g)$ is equivalent to a direct sum of $m$ copies of $\sigma$.   
We can rewrite Eq.~\eqref{eq:RepInterwiner} in tensor graphical form as %
\begin{equation}\label{eq:RepInterwinerTN}
	\begin{tikzpicture}[baseline={([yshift=-1.4ex]current bounding box.center)}, scale=.75]
		\umatrix{0}{0}{V}{a}
		\indexflowout{-1.4}{0}{-1}{}
		\indexflowoutV{0}{1.4}{1}{}
		\indexflowoutV{0}{0}{1}{}
		\indexflowin{0}{0}{1}{}
		\indexflowin{-1.4}{0}{1}{}
		\circleTensor{-1.4 }{0}{\small \sigma(g)}
		\circleTensor{0}{1.4}{\small \psi(g)}
	\end{tikzpicture}=
	\begin{tikzpicture}[baseline={([yshift=-.6ex]current bounding box.center)}, scale=.75]
		\umatrix{0}{0}{V}{a}
		\indexflowout{0}{0}{-1}{}
		\indexflowoutV{0}{0}{1}{}
		\indexflowin{0}{0}{1}{}
		\indexflowin{1.4}{0}{1}{}
		\circleTensor{1.4}{0}{\small \sigma(g)}
	\end{tikzpicture}.
\end{equation} 
The tensor $V$ satisfies the following algebraic relation
\begin{equation}\label{eq:RXweakequiv}
	\begin{tikzpicture}[baseline={([yshift=.4ex]current bounding box.center)}, scale=.8]
		\umatrix{0}{0}{V}{}
		\umatrix{1}{0}{V}{}
		\Rmatrix{0.5}{-1}{R}
	\end{tikzpicture}= \begin{tikzpicture}[baseline={([yshift=.4ex]current bounding box.center)}, scale=.8]
		\pimatrix{0.5}{1}
		\umatrix{0}{0}{V}{}
		\umatrix{1}{0}{V}{}
	\end{tikzpicture}, 
\end{equation}
where the $R$-matrix is the same as in Eq.~\eqref{eqApp:seth-R} but without the minus sign~(it is clear that the minus sign does not affect the winning strategy). 

To win the game, the players prepare the physical system $(\hat{H},\ket{G})$ on a topologically trivial manifold,  where 
\begin{equation}
\hat{H}=\hat{H}_0+\hat{V}^{\sigma}_o+\hat{V}^{\bar{\sigma}}_s,
\end{equation}
and $\hat{V}^{\sigma}_o$ and $\hat{V}^{\bar{\sigma}}_s$ are the trapping potentials for $\sigma$ and $\bar{\sigma}$, respectively, as defined below Eq.~\eqref{eq:trappingpotential}, %
and
\begin{equation}
\ket{G}=\adjincludegraphics[height=4.5ex,valign=c]{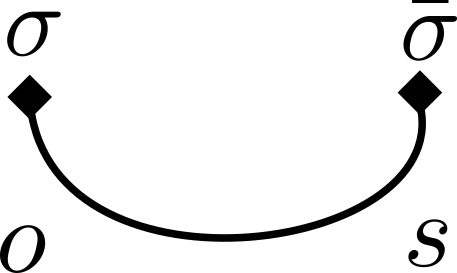}\ket{G_0}
\end{equation}	
is the unique, gapped ground state of $\hat{H}$. 

One can create a quasiparticle $\psi$ above the defected ground state $\ket{G}$ by applying an open Wilson line labeled by $\psi$: 
\begin{equation}\label{eq:Gpsi-GT}
	\ket{G;i^a}=\adjincludegraphics[height=7ex,valign=c]{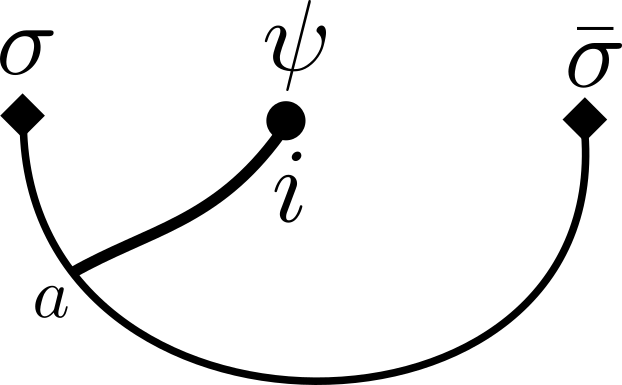}\ket{G_0}, 
\end{equation}
where the trivalent vertex denotes the intertwiner $V$ defined in Eq.~\eqref{eq:RepInterwinerTN}. Importantly, the location of the trivalent vertex can be continuously moved along the $\sigma$-line with out changing the overall quantum state:
\begin{equation}\label{eq:IntertwinerMove-abs}
	\adjincludegraphics[height=7ex,valign=c]{Figures/ParaExchange/PsiiaG1.png}\ket{G_0}
	=\adjincludegraphics[height=7ex,valign=c]{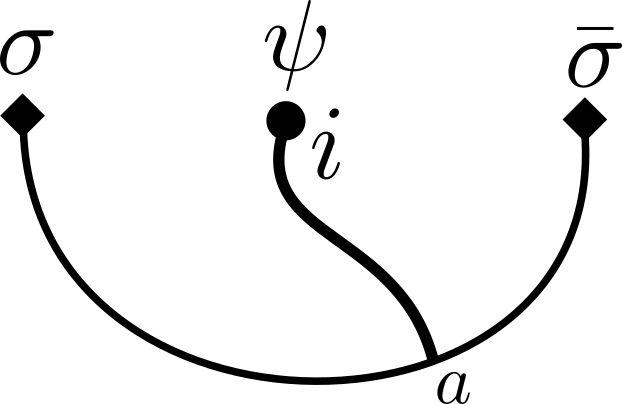}\ket{G_0}.
\end{equation}
This follows from the following elementary move at the microscopic level:
\begin{equation}\label{eq:IntertwinerMove}
	\adjincludegraphics[height=18ex,valign=c]{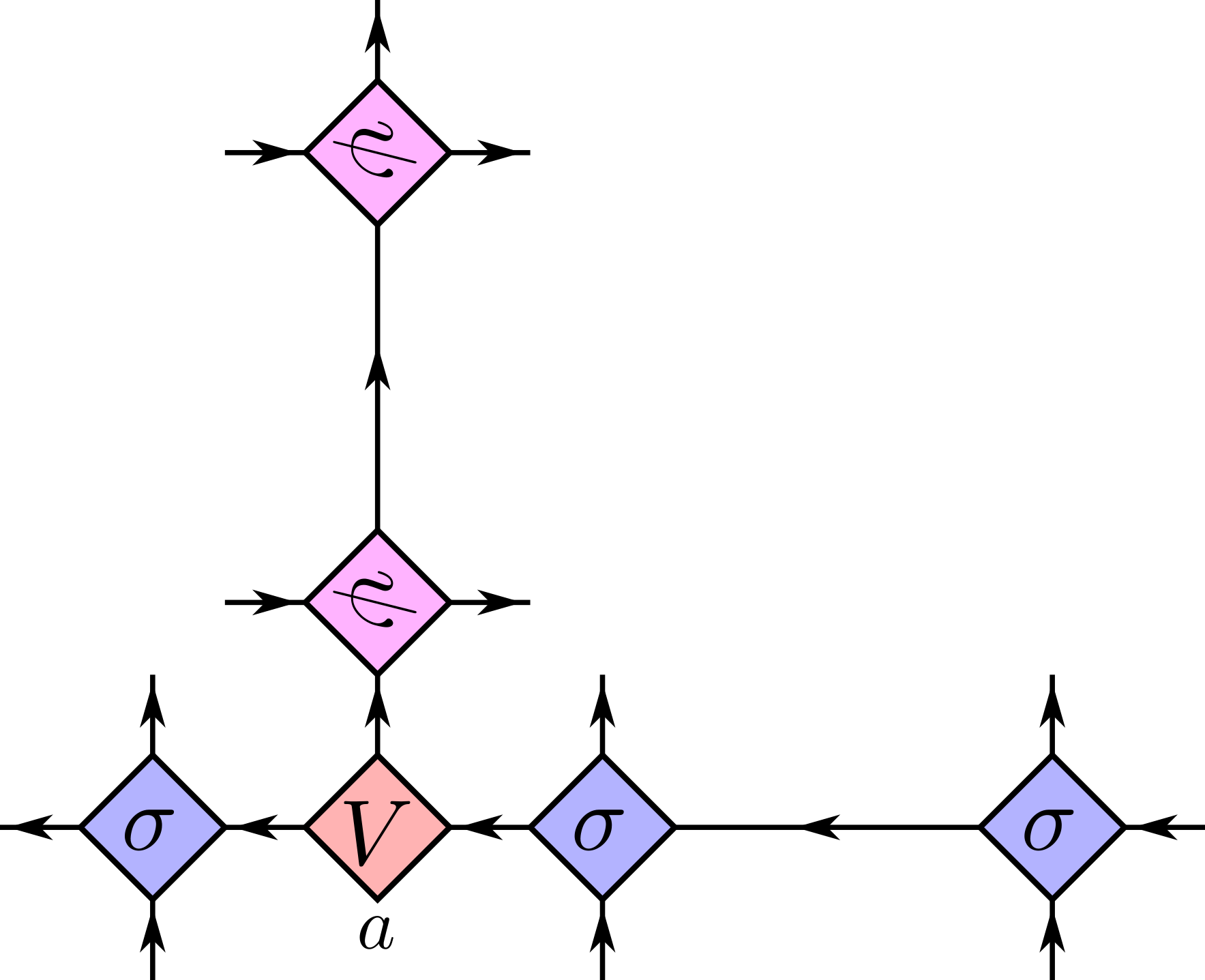}
	=\adjincludegraphics[height=18ex,valign=c]{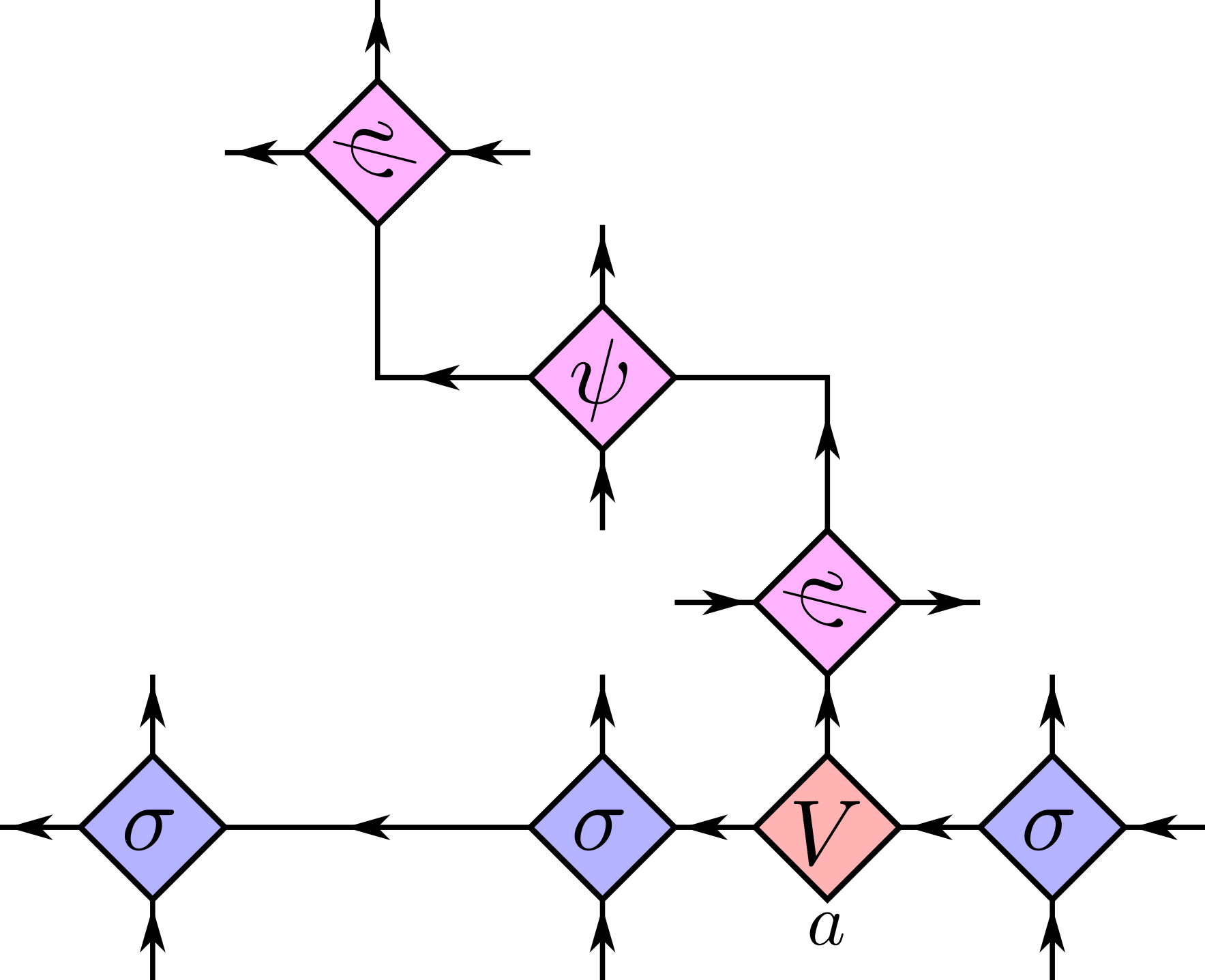}.
\end{equation}
This property implies that the index $a$ in $\ket{G,i^a}$ labels a topological degree of freedom that cannot be accessed by any local operations when $\psi,\sigma,\bar{\sigma}$ are far-separated.  

In the following we show that the quasiparticle $\psi$ in this model exhibits the
universal properties of emergent $R$-paraparticles presented in Eqs.~(\ref{eq:TPdegenerate}-\ref{eq:localmeasurementatcorner}):\\ 
(1) 
The basis states for the $n$-particle subspace $\Hil_{i_1i_2 \ldots i_n}$ are defined as %
\begin{equation}\label{eq:basis_n_particle_space-GT}
	\ket{G;i_1^{a_1} i_2^{a_2} \ldots i_n^{a_n}}=\adjincludegraphics[height=10ex,valign=c]{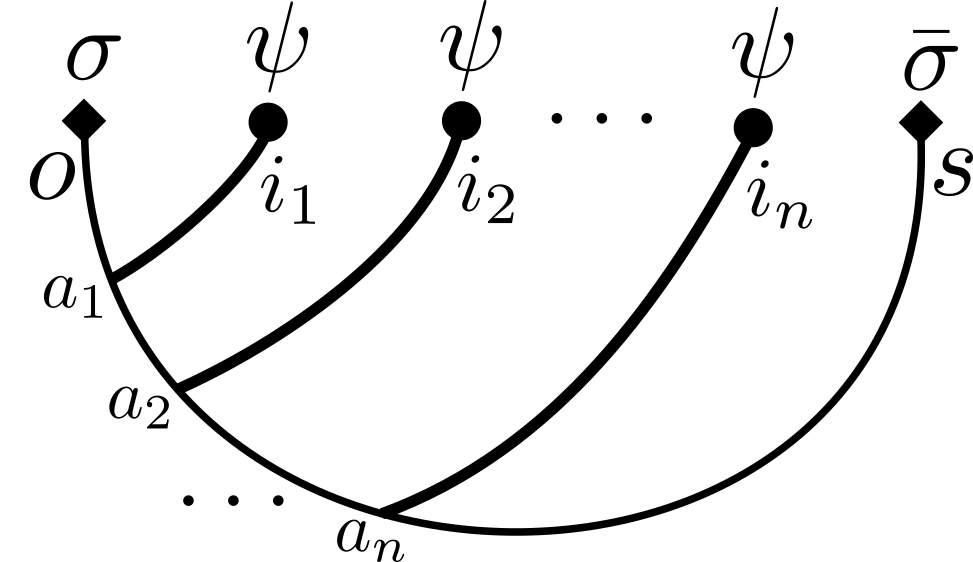}\ket{G_0}.
\end{equation}
Indeed, $\Hil_{i_1i_2 \ldots i_n}$ is isomorphic to the multiparticle fusion space $V^{\sigma\psi\psi\ldots\psi\bar{\sigma}}_{I}$.\\
(2) Topological degeneracy of $\Hil_{i_1i_2 \ldots i_n}$ directly follows from the fact that the indices $a_1,a_2,\ldots,a_n$ are not locally accessible, as we mentioned below Eq.~\eqref{eq:Gpsi-GT}.  \\
(3) 
It is known that in deconfined gauge theories, all quasiparticles can be moved by applying Wilson line operators. A  useful fact to note here is that the tensor defined in Eq.~\eqref{eq:PsiTensor} is actually a dual unitary tensor~\cite{Prosen2019DU}. In this way, one can use the Wilson line operator in Eq.~\eqref{eq:PsiLine} to construct a unitary ladder operator that moves the quasiparticle $\psi$ from one location to another. \\
(4) Exchange statistics in Eq.~\eqref{eq:exchangestatR} follows from
\begin{equation}\label{eq:paraexchange}
	\adjincludegraphics[height=8ex,valign=c]{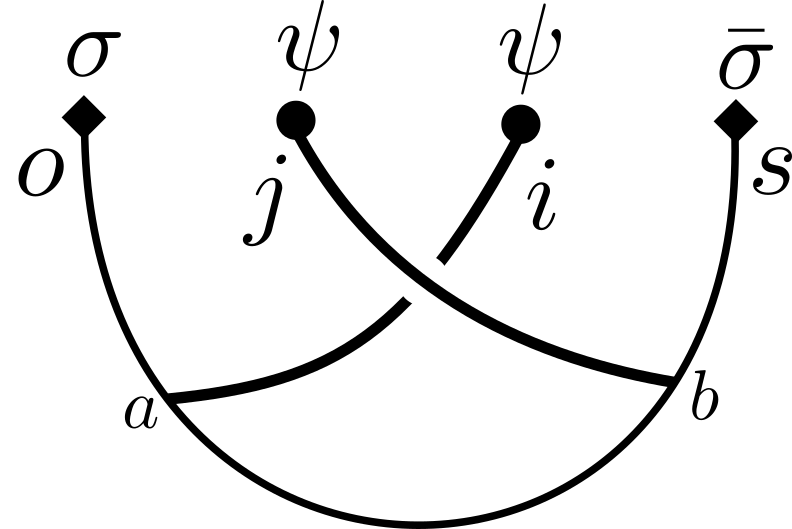}
	=\sum_{b',a'}R^{b'a'}_{ab}~\adjincludegraphics[height=8ex,valign=c]{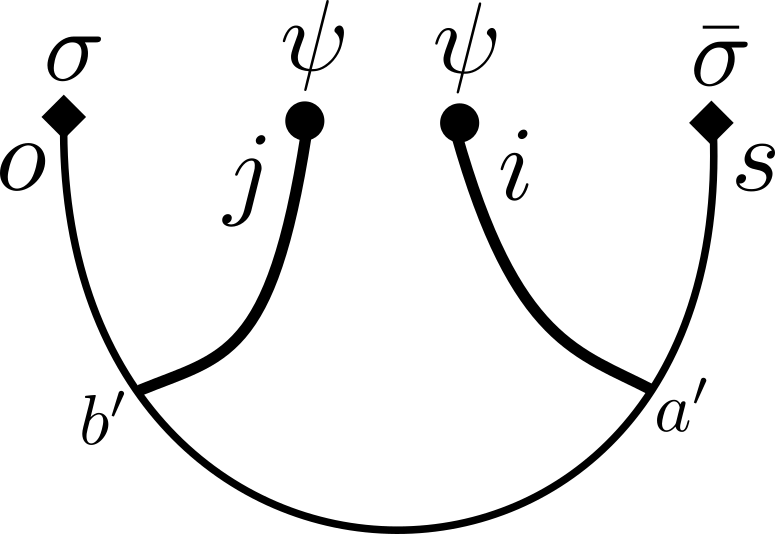},
\end{equation}
where we use Eq.~\eqref{eq:RXweakequiv} to swap the two intertwiners.\\
(5) 
The local unitary operators $\hat{U}_{\oA,a}$, $\hat{U}'_{\oB,a}$ in Eq.~\eqref{eq:localcreationatcorner} can be constructed using elementary Wilson line operators. For example, one can construct $\hat{U}_{\oA,a}$ such that
\begin{equation}\label{eq:localcreation_U-GT}
	\hat{U}_{\oA,a}\adjincludegraphics[height=10ex,valign=c]{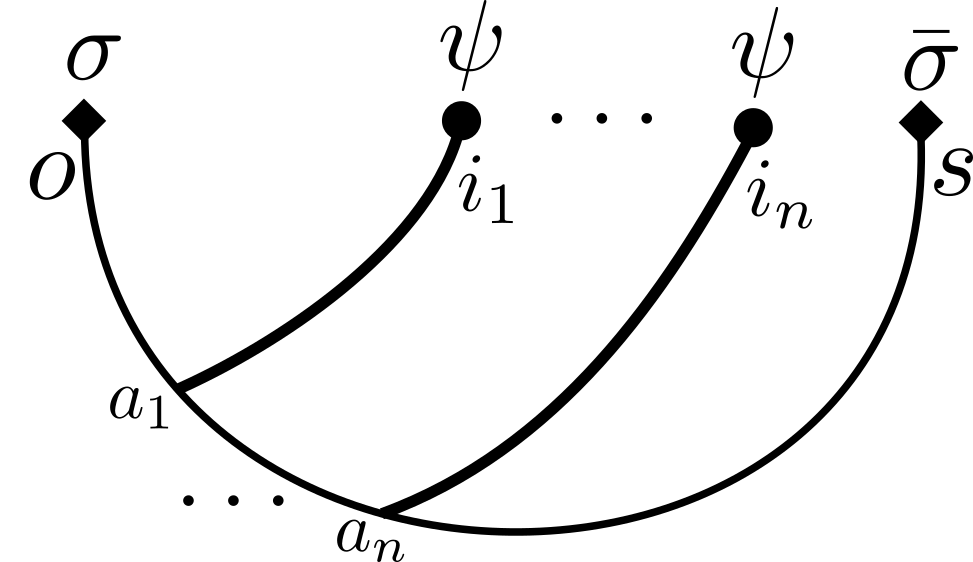}\ket{G_0}=
	\adjincludegraphics[height=10ex,valign=c]{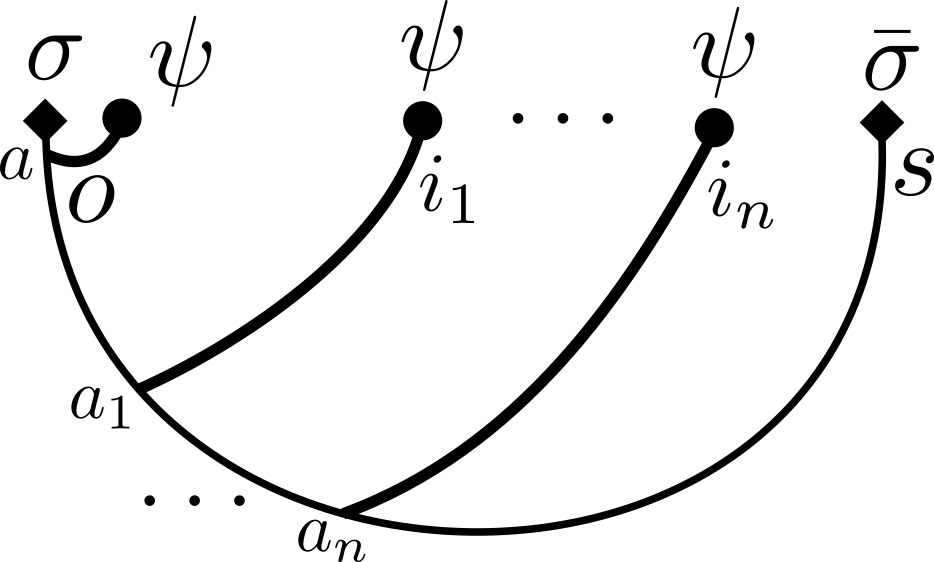}\ket{G_0}.
\end{equation}  
(6) 
When a $\psi$ is close to $\sigma$, one can make local measurements in the fusion space $V_\sigma^{\sigma\psi}$:
\begin{equation}\label{eq:def_color_operator-GT}
	\hat{O}_\oA\adjincludegraphics[height=10ex,valign=c]{Figures/ParaExchange/UoaGR.png}\ket{G_0}
	=a~\adjincludegraphics[height=10ex,valign=c]{Figures/ParaExchange/UoaGR.png}\ket{G_0}.
\end{equation}
We do not give an explicit form for the local operators in (5) and (6), as microscopic details are not important for the winning strategy, and knowing their existence is enough. 

Given that $\psi$ in this model satisfies all the universal properties (1)-(6) with a nontrivial $R$-matrix, we conclude that the system $(\hat{H},\ket{G})$ can win the game, using the same strategy introduced in the main text. We emphasize that with this system  $(\hat{H},\ket{G})$, the players can win the game without violating any rules. Importantly, the rules of the game require that, when the circles move away from the starting points $\oA,\oB$, the players do not leave any \textit{extra} excitations at $\oA,\oB$ \textit{relative} to the ground state $\ket{G}$. The topological defects $\sigma,\bar{\sigma}$ are excitations above the translationally invariant state $\ket{G_0}$, but they are already present in $\ket{G}$, and they sit still at $\oA,\oB$ throughout the game process, so they are considered as part of the background in $\ket{G}$, not extra excitations created by the players after the game has started. As described in the game protocol, during the game, the Referees monitor the system by measuring the local Hamiltonians in $\hat{H}_{}$ everywhere beyond the circle areas, so with this strategy they will not detect any extra excitations at $\oA, \oB$ relative to the ground state $\ket{G}$. 
\section{The anti-anyon twist in 2D}\label{SI:AAtwist}
\begin{figure}
\begin{subfigure}[t]{.45\linewidth}
\centering\includegraphics[width=.8\linewidth]{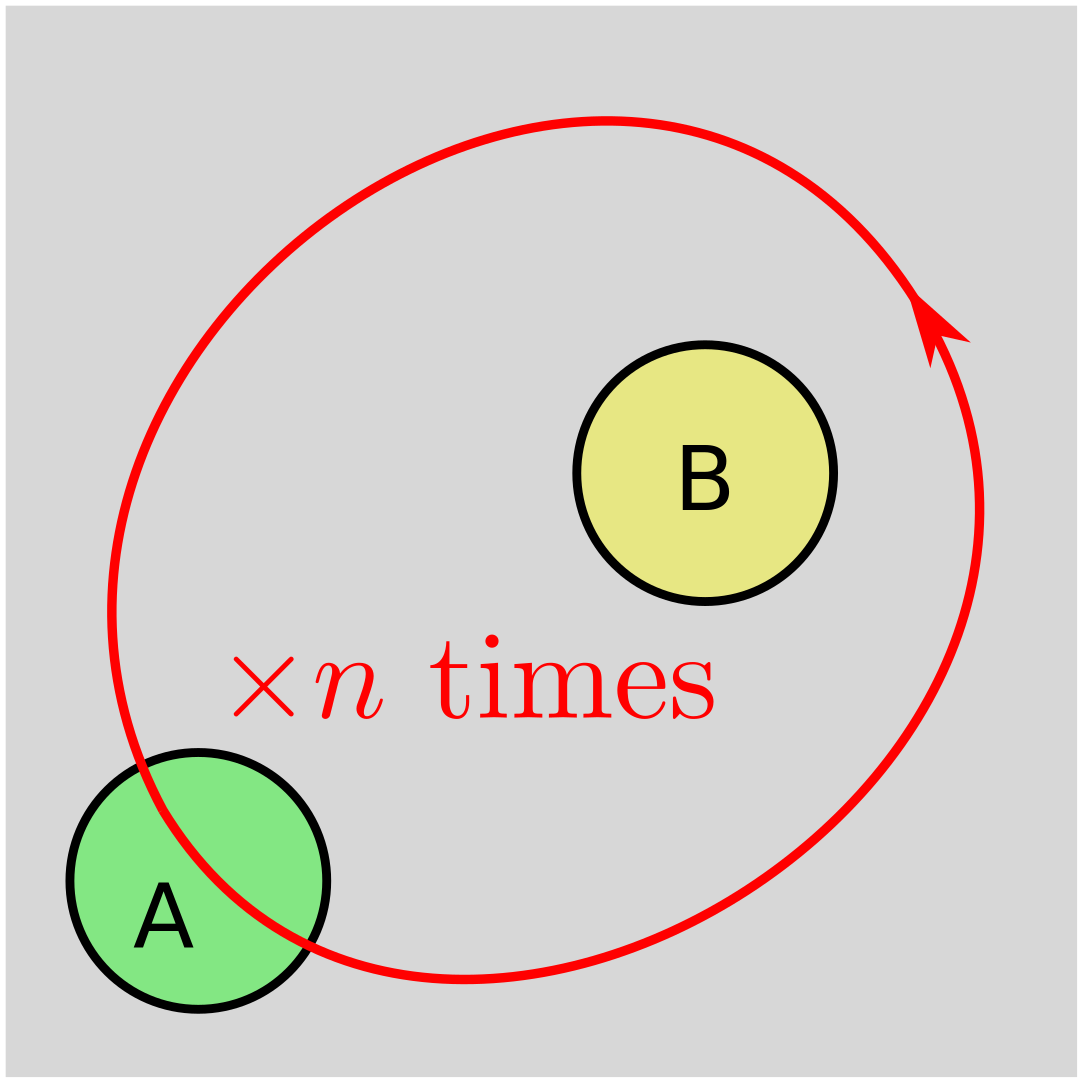}
\caption{\label{fig:AAtwist-protocol} Game protocol.}
\end{subfigure}
\begin{subfigure}[t]{.45\linewidth}
\includegraphics[width=.8\linewidth]{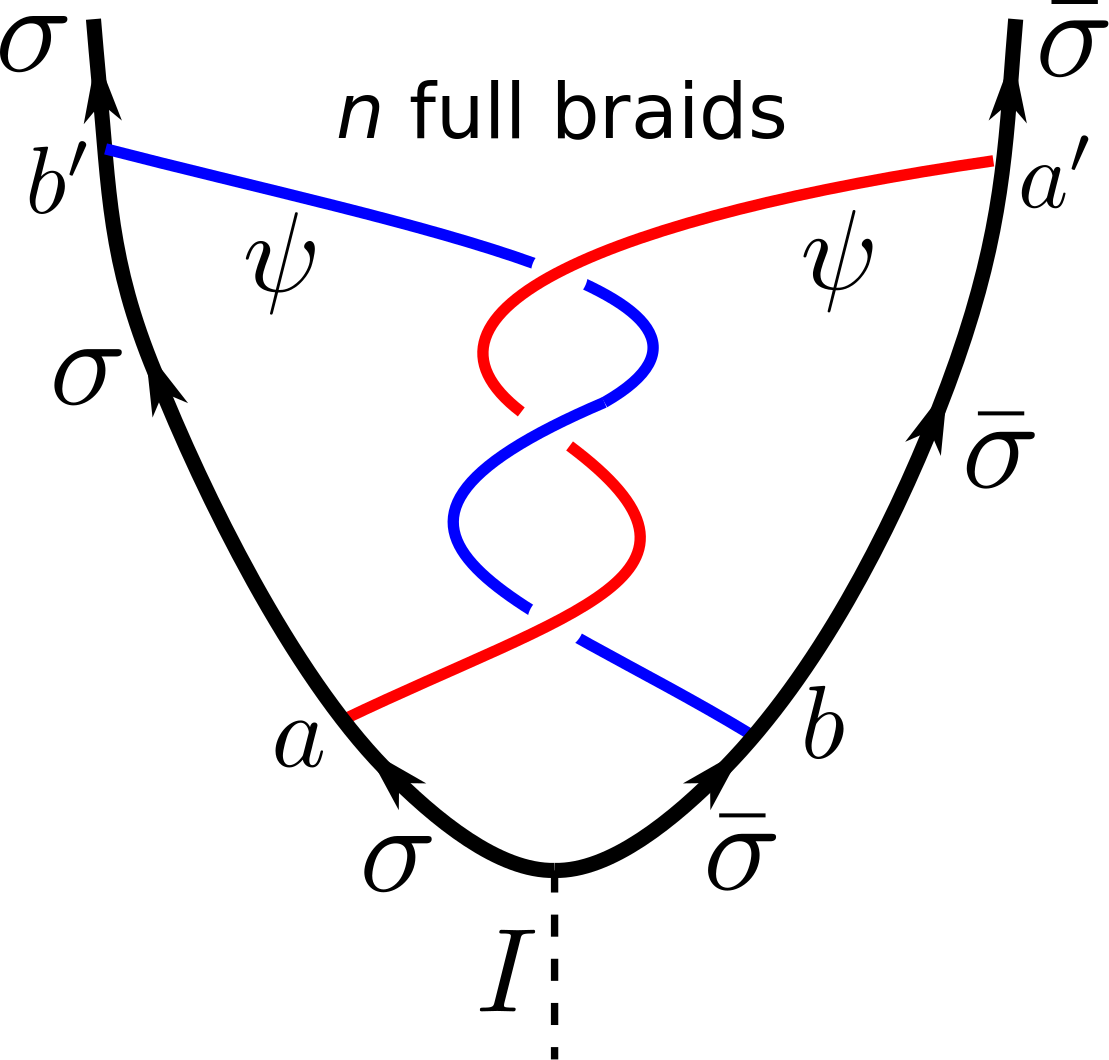}
\caption{\label{fig:AAtwist-SFC} Diagrammatic description. }
\end{subfigure}
\caption{\label{fig:AAtwist} The anti-anyon twist in 2D. }
\end{figure}
As we mentioned in the main text, it is possible that certain types of non-Abelian anyons can also win the 2D version of the game~(while the 3D version can only be won using paraparticles, as the mobility of anyons are restricted to 2D). %
If we want to physically distinguish  paraparticles even in the 2D  version of the game, we can add some twists to the game protocol to prevent non-Abelian anyons from winning. 
In the following, we make a preliminary attempt in this direction, and present a simple example of anti-anyon twist, which %
prevents a class of non-Abelian anyons from success. More complicated anti-anyon twists that rule out all anyon-based winning strategies %
will be presented in a future work~\cite{wang2025secret}.

The basic idea of the anti-anyon twist is to exploit the fundamental distinction between paraparticles and non-Abelian anyons: for the former, a full braid of one particle around another does not change the quantum state of the system~(since by definition $R^2=\mathds{1}$ for paraparticles), while full braids between non-Abelian anyons are generally nontrivial. 
This motivates the idea of adding random full braids to the game protocol to scramble anyon-based strategies. 
Specifically, let us suppose that in the middle of the game process, when both players are  far away from the boundaries~(see Fig.~\ref{fig:AAtwist-protocol}), the Referees have the discretion to temporarily halt the movement of circle B, and slowly move the circle A around circle B $n$ times, where $n\in\mathbb{Z}$ is chosen randomly by the Referees.   
Note that during the entire process Alice is still obliged to move whatever excitations inside the circle to follow the circle movement, and importantly, Bob does not know the value of $n$~(because each player is confined in a separate room and cannot know what the other circle is doing). After this is done, the game proceed as in the original game protocol. 

We now analyze the effect of this anti-anyon twist on the winning strategy presented in the main text. Generalizing Eq.~\eqref{eq:Psi3} in the main text, the time evolution of the system is now described by the braided fusion diagram shown in Fig.~\ref{fig:AAtwist-SFC}, which takes into account the extra $n$ full braids added to the exchange process. %
Consequently, the $R$ in the RHS of Eq.~\eqref{eq:Psi3} now becomes $R^{2n+1}$. The reduced density matrix describing Bob's measurement result is
\begin{eqnarray}\label{eq:reducedDM_B}
\rho_B&=&\frac{1}{n}\sum_n\Tr_A[\ket{\Psi_n}\bra{\Psi_n}],\nonumber\\
\ket{\Psi_n}&=&(R^{2n+1})\ket{a}_A\otimes\ket{b}_B,
\end{eqnarray}
where in the second line $R^{2n+1}:V_A\otimes V_B\to V_B\otimes V_A$ is considered as a unitary operator and $V_A$~($V_B$) is the internal space of Alice's~(Bob's) particle, and $\Tr_A$ is the partial trace over $V_A$. Importantly, since Bob has no knowledge about the number $n$, we average over $n$ to obtain $\rho_B$. 
As we can see, for non-Abelian anyons,  $R^2\neq \mathds{1}$ will randomly scramble Bob's measurement result, and $\rho_B$ will contain less information about $a$. If $R^2$ is sufficiently complicated to completely remove the dependence of $\rho_B$ on $a$, then Bob cannot obtain any information about $a$ and will therefore fail the challenge~\footnote{Here Alice may still able to calculate Bob's number correctly by carefully analyzing her circle movements to figure out the unknown number $n$. But we can easily prevent this by adding another round of the above twist with the role of Alice and Bob exchanged, in which circle B moves around circle A $k$ times.  Then the $R^{2n+1}$ in Eq.~\eqref{eq:reducedDM_B} will be replaced by $R^{2n+2k+1}$, so one needs to know the value of $n+k$ to correctly calculate each other's number. But Alice does not know $k$, while Bob does not know $n$, so both of them will fail. }. 

Therefore, by adding random full braids to the game protocol, we can scramble anyon-based strategies without complicating the original strategy using paraparticles. As we mentioned, the above simple twist can only prevent a small class of non-Abelian anyons for which $R^2$ is sufficiently complicated to completely scramble $\rho_B$. In Ref.~\cite{wang2025secret}, we introduce more powerful~(yet more complicated) anti-anyon twists including multiplayer generalizations of the game protocol and an anyon interference test, towards eventually preventing all anyon-based strategies.

\section{Robustness against noise and eavesdropping}\label{sec:noise_robust}
In the main text, we claimed that the winning strategy using emergent paraparticles is robust against local noise and eavesdropping. Below we explain the underlying assumptions behind this robustness in greater detail.

The claim that the winning strategy is robust against noise is based on the following assumptions:\\
(1). \textit{All noise in the system are local, i.e. noises are modeled by local interactions between system and environment.} This is a very common assumption used in the analysis of stability of topological order~\cite{hastings2005quasiadiabatic1} and robustness of topological quantum computation~\cite{kitaev2003fault,Nayak2008NAAnyons}. In general, the information stored in the topological degrees of freedom~(the topologically protected space) can only be accessed and modified when some topological excitations are either close to each other or close to some topological defects--in such a situation one can locally access and modify the information via fusion, and this is the only time when the information is susceptible to local noise and eavesdropping. In our case, since we assume that two circles are always far away from each other, local noise can affect the topologically stored information only when both paraparticles are close to the boundary, i.e. at time around $t=0$ and $t=T$. We now further assume that\\
(2). \textit{The noise rate is upper bounded by a constant that is independent of the system size};

Since the susceptibility to local noise decays exponentially in the distance between the paraparticles and the boundary, the total influence of the noise to the protected information is upper bounded by a constant independent of the system size. This means that in the winning strategy using paraparticles, even in the presence of local noise, the players can still transfer an amount of information to each other independent of system size.

However, even though local noises cannot significantly damage the topologically protected information, they may still invalidate the winning strategy in a different way, by accidentally breaking the rules of the game. 
Imagine that in the middle of the game, a local noise~(modeled by a random local unitary operation $\hat{U}_{\text{noise}}$) creates a local excitation somewhere in the system, say at the point $\oA$. 
Then, according to the game protocol, the next time when the Referees check the ground state projector at $\oA$, they will detect a local excitation there, and wrongly conclude that Alice has cheated by leaving a local excitation at $\oA$ and therefore declare that challenge fails, although this is not caused by the players at all. To resolve this issue, we can slightly modify the game protocol as follows. Whenever the Referee detects a local excitation anywhere in the system beyond the circle areas, instead of declaring that the challenge fails, the Referees will first try to remove the excitation using local unitary operations at the vicinity of the point where the excitation is detected. If they are able to completely remove the excitation using local operations, then the game continues. Only if they cannot remove the excitation using any kind of local operation~\footnote{
This can happen, for example, if the system has a fusion rule of the form $\rho\times \psi=\sigma+\ldots$, and at $t=0$, Alice creates a particle $\psi$ in her circle, transforming the defect type $\sigma$ into a different type $\rho$. This is considered cheating~(as it leaves locally-detectable information behind) and is prohibited by the game protocol. In this case, after Alice's circle moves away from $\oA$, the Referees will not be able to transform $\rho$ back to $\sigma$ using only local operations at $\oA$, and will therefore declare that the challenge fails.
} will they declare that the challenge fails~\footnote{
The readers may worry that the Referees may not have enough motivation to help the players correct the local errors caused by environmental noise. To address this concern, we can introduce another player into the game, called the corrector~(denoted by C), who plays in the same team with Alice and Bob to win the challenge. The corrector is allowed to discuss the winning strategy with Alice and Bob before the game begins. During the game, C is also isolated in a separate room  similar to Alice and Bob. %
Whenever the Referee detects a local excitation anywhere in the system beyond the circle areas, they first asks C to remove the excitation using local operations restricted to a circle area centered around the excitation~(of course, we always guarantee that C's circle never overlaps with those of Alice and Bob). If C successfully removes the excitation~(as the Referees verify by measuring local ground state projectors), then the game continues; otherwise, the challenge fails.  
}.  
With these relaxed game rules, local noise cannot accidentally terminate the game, and the Referees can still prevent the players from using the trivial way of communication by leaving the information locally at $\oA$ and $\oB$, since any information stored locally will be erased by the Referees.

\section{Categorical analysis of the winning strategy}\label{sec:Cat_analysis}
Recall that the main objective of this work is to propose a challenge game that demonstrates the \textit{unique} physical feature of paraparticles, as we claimed at the very beginning. To establish this fact, we need to show that a physical system can win the challenge game \textit{if and only if} it hosts emergent paraparticles. The ``if'' part has already been established in the main text, where we demonstrated a winning strategy for physical systems hosting emergent paraparticles~[in the sense of satisfying the properties~(1-6) in the main text with a nontrivial $R$-matrix]. The ``only if'' part turns out to be much harder to establish on rigorous grounds. In the main text, we presented an intuitive argument that non-trivial exchange statistics is needed to win this challenge. 
We now try to promote this intuitive argument into a more careful analysis using the categorical description of topological phases~\cite{kitaev2006anyons}, and show that under additional assumptions, only physical systems hosting emergent paraparticles can win the 3D version of the challenge. Here we only sketch the key steps, and details are presented in Ref.~\cite{wang2025secret}. For simplicity, we focus on the version of the challenge game without any boundary. 

The first important observation is that the set of physical systems that can pass the challenge is invariant under local unitary transformations~(LUTs)~\cite{Chen2010LUT}. More precisely, suppose a physical system described by a locally-interacting Hamiltonian $\hat{H}$ can pass the challenge, and let $\ket{G}$ be its unique, gapped, and frustration-free ground state. The winning strategy can be described by the sequence of physical operations performed by the players,  encoded in the functions $\hat{U}_A(t),\hat{O}_A(t), \hat{U}_B(t),\hat{O}_B(t)$ for $0\leq t\leq T$, where  $\hat{U}_A(t)$ and $\hat{O}_A(t)$ are the local unitary operation and measurement performed by Alice at time $t$, and similarly for $\hat{U}_B(t)$ and $\hat{O}_B(t)$. Now let $\hat{U}$ let any LUT that can be represented as a finite depth unitary circuit~\footnote{Note that for simplicity, here we only consider a subclass of LUTs that have strictly finite depth. But the arguments here can be straightforwardly extended to the more general class of LUTs if we relax the frustration-free requirement as we mentioned in Sec.~\ref{sec:relax_FF}. }. It is then clear that the physical system described by the transformed Hamiltonian $\hat{H}'=\hat{U}\hat{H}\hat{U}^\dagger$ can also pass the challenge. Indeed, since  $\hat{U}$ is a finite depth local unitary circuit, $\hat{H}'$ is also locally-interacting, and its ground state $\ket{G'}=\hat{U}\ket{G}$ is also unique, gapped, and frustration-free, thereby qualifying the requirements of the game. It is then clear that the players can simply use the transformed operations $\hat{U}\hat{U}_A(t)\hat{U}^\dagger,\hat{U}\hat{O}_A(t)\hat{U}^\dagger, \hat{U}\hat{U}_B(t)\hat{U}^\dagger,\hat{U}\hat{O}_B(t)\hat{U}^\dagger$ to win the game~\footnote{One subtlety here is that these transformed operations may be supported on a larger area due to $\hat{U}$ being a finite depth local  unitary circuit, so the players may need a larger $r_0$ for the radius of the two circles~(or spheres in 3D). %
But this is not a problem given that the game protocol allows the players to choose $r_0$. 
}. 

The above fact implies that whether a physical system can pass the challenge depends only on its underlying topological order, which is an equivalence class of gapped ground states under the equivalence relation defined by LUTs~\cite{Chen2010LUT}.
This allows us to forget about microscopic details and focus on universal topological properties. 

As we argued in the main text, since the game protocol allows the Referees to choose the path after the players submit the system, the quasiparticle excitations in the two spheres need to have unrestricted mobility so that the players can manage to keep the excitations within the spheres throughout the game. It is widely believed that the universal properties of quasiparticles with unrestricted mobility that can emerge in 3D gapped phases are described by SFCs~\cite{LanKongWen3DAB,LanWen3DEF}, and we make this assumption in the following analysis. Actually, for technical reasons, here we make a stronger assumption that \textit{all} topological excitations of the system are described by an SFC, including not only the quasiparticles in the two spheres, but also the two topological defects $\sigma,\sigma'$ that sit stationary at $\oA$ and $\oB$ throughout the game. While it is clear from prior arguments that $\psi_1$ and $\psi_2$ must have unrestricted mobility, one may perceive a winning strategy in which $\sigma$ and $\sigma'$ have restricted mobility, such as fractons~\cite{haah2011local,vijay2016fracton1,nandkishore2019fractons}. This case can be interesting to study in its own right, but it is way beyond the scope of this paper.

From the requirement that the ground state $\ket{G}$ must be unique and gapped, it is straightforward to argue that, on a closed manifold without boundary and defects beyond the two points $\oA$ and $\oB$, we must have $\sigma'=\bar{\sigma}$~(the antiparticle of $\sigma$), and in addition, $\sigma$ must be simple. 

Now let $I_T$ denote the time interval during the game when both spheres do not contain either $\oA$ or $\oB$.
Let $\psi$ be the total fusion channel of all the excitations within sphere A, and  $\psi'$ is defined similarly for sphere B. It is clear that both $\psi$ and $\psi'$ cannot be changed during $I_T$ since the players are only allowed to perform local operations within their respective spheres.  In the last part of the main text, we described a winning strategy using SFC language, described by the fusion diagram in Fig.~\ref{eq:SFCdescriptiongame}, where we have $\psi'=\psi$, i.e., Alice and Bob use the same type of particles to win the game. The question is, is $\psi'=\psi$ necessary? In the next paper we present an example in which Alice and Bob use different types of particles $\psi'\neq \psi$ with non-trivial mutual parastatistics to win the challenge. 
While we can consider mutual parastatistics as a generalization of parastatistics, if we want to single out self-parastatistics~[as described by properties (1-6) in the main text], we can add the so-called identical particle test introduced in Ref.~\cite{wang2025secret} to prevent mutual parastatistics from winning. By requiring the players to also pass the identical particle test, one can show that we must have  $\psi'=\psi$ and $\psi$ must be simple. [Actually, even without the identical particle test, one can still argue that both $\psi'$ and $\psi$ must be simple for the strategy to be robust against noise and eavesdropping. Then the main conclusion still holds up to the possibility of mutual parastatistics.]

Now consider the moment at early game when both spheres are just about to leave the points $\oA$ and $\oB$. Since the particle type of the topological defects at $\oA$ and $\oB$ are required to be always equal to $\sigma$ and $\bar{\sigma}$, respectively~(because the Referees constantly check the local ground state projectors at these two points during $I_T$), in order to guarantee that Alice can locally create a $\psi$ in the vicinity of $\sigma$ and Bob can locally create a $\psi$ in the vicinity of $\bar{\sigma}$, the SFC must have fusion rules of the form 
\begin{equation}\label{eq:sigmapsifusion-gen}
\sigma \times \psi=m~\sigma+\rho,\quad   \psi\times \bar{\sigma}=m~\bar{\sigma}+\rho',
\end{equation}
for some integer $m\geq 1$, where $\rho\in \mathcal{C}$ collectively denote the sum of all other possible particle types contained in the fusion product $\sigma \times \psi$, and similarly for $\rho'$. 
Now consider the moment close to game ending when $\oA$ is inside Bob's sphere. 
After Bob performs some measurement the total fusion channel in Bob's sphere becomes a fusion product of $\sigma$ and $\psi$. 
If $\rho\neq 0$, then there is the danger that $\sigma$ and $\psi$ may fuse into $\rho$ which will fail the challenge, because according to the game protocol, the Referees will perform another round of local ground state condition checks even after both spheres disappear. If Bob's measurement gives $\rho$, then there is no way for him to change $\rho$ back to $\sigma$ using any local operations. Therefore we need $\rho=0$~(there may actually be some small exceptions to this argument which we will treat more carefully in the next paper), and similarly $\rho'=0$. 

Thus we arrive at the fusion rule in Eq.~\eqref{eq:sigmapsifusion} of the main text, and the time evolution of the system is described by the fusion diagram in Eq.~\eqref{eq:SFCdescriptiongame}. It is then straightforward to show that $R$ must be nontrivial to enable information transfer between Alice and Bob. 
More precisely, let $\rho_A(t)$ and $\rho_B(t)$ be the local reduced density matrix within the spheres A and B at $t$, respectively. Then one can show that $\{U_B(t)\}_{0\leq t\leq T}$ can influence $\{\rho_A(t)\}_{0\leq t\leq T}$
if and only if $R$ is nontrivial~(and similarly if one swaps $A\leftrightarrow B$)~\cite{wang2025secret}. %
The completes the SFC description of the winning strategy. 
Finally, one can show that the $R$-matrix defined by Eq.~\eqref{eq:SFCdescriptiongame} must satisfy the Yang-Baxter equation in Eq.~\eqref{eq:YBE}, and 
the particle $\psi$ satisfies all the properties (1-6) in the main text, i.e., $\psi$ is an emergent paraparticle~\cite{wang2025secret}.  

We finally mention that all the analysis above also apply to the 2D case, provided that one replaces SFCs  by braided fusion categories~(or modular tensor categories), which describe universal  properties of topological quasiparticles in 2D gapped phases of matter. The conclusion is similar: a 2D topological phase described by a modular tensor category $\calC$ can win the game if $\calC$ has a fusion rule of the form in Eq.~\eqref{eq:sigmapsifusion} of the main text such that the $R$-matrix defined by Eq.~\eqref{eq:SFCdescriptiongame} is nontrivial~(i.e., not of the trivial product form $R^{b'a'}_{ab}=p_{a'a}q_{b'b}$). The only difference is that in 2D, $R$ produced by a braided fusion category need not be involutive: if $R^2\neq \mathds{1}$, $\psi$ is a non-Abelian anyon, while if $R^2= \mathds{1}$, $\psi$ is a paraparticle. Therefore, both $R$-paraparticles and a special class of non-Abelian anyons can win the 2D game, while the 3D game can only be won by $R$-paraparticles, since anyons do not exist in 3D.  %
\end{document}